\documentclass[12pt]{article}
\usepackage{fullpage}
\usepackage{setspace}
\usepackage{epsf}
\usepackage{natbib}
\usepackage{epsfig}
\usepackage{amsthm}
\usepackage{color}
\usepackage{amsmath}
\usepackage{amssymb}

\title{
Gaussian Processes and Limiting Linear Models
}

\author{
  Robert B. Gramacy and Herbert K. H. Lee\thanks{\noindent Robert
    Gramacy is a Lecturer in the Statistical Laboratory, University of
    Cambridge, UK (Email: bobby@statslab.cam.ac.uk) and Herbert Lee is
    a Professor in the Department of Applied Mathematics and
    Statistics, University of California, Santa Cruz, CA 95064 (Email:
    herbie@ams.ucsc.edu).  This work was partially supported by NASA
    awards 08008-002-011-000 and SC 2003028 NAS2-03144, Sandia
    National Labs grant 496420, and National Science Foundation grants
    DMS 0233710 and 0504851.}
\date{\mbox{}\vspace*{-2.5cm}\mbox{}}
}

\begin{document}

\doublespacing

\newcommand{\bm}[1]{\mbox{\boldmath $#1$}}
\newcommand{\mb}[1]{\mathbf{#1}}

\maketitle

\begin{abstract}
  Gaussian processes retain the linear model either as a special case,
  or in the limit.  We show how this relationship can be exploited
  when the data are at least partially linear.  However from the
  perspective of the Bayesian posterior, the Gaussian processes which
  encode the linear model either have probability of nearly zero or
  are otherwise unattainable without the explicit construction of a
  prior with the limiting linear model in mind.  We develop such a
  prior, and show that its practical benefits extend well beyond the
  computational and conceptual simplicity of the linear model.  For
  example, linearity can be extracted on a per-dimension basis, or can
  be combined with treed partition models to yield a highly efficient
  nonstationary model.  Our approach is demonstrated on synthetic and
  real datasets of varying linearity and dimensionality.  

\end{abstract}

\section{Introduction}

The Gaussian Process (GP) is a common model for fitting arbitrary
functions or surfaces, because of its nonparametric flexibility
\citep{cressie:1991}.  Such models and their extensions are used in a
wide variety of applications, such as computer experiments
\citep{kennedy:ohagan:2001,sant:will:notz:2003}, environmental
modeling \citep{gill:nych:2005,cald:2007}, and geology
\citep{chil:delf:1999}.  The modeling flexibility of GPs comes with
large computational requirements and complexities.  Sometimes a GP
will produce a very smooth fit, i.e., one that appears
linear\footnote{We use ``smooth'' in this colloquial sense throughout
  to mean nearly linear, or the opposite of ``wavy'', as opposed to
  the more technical ``infinitely differentiable'' sense.}.  In these
cases, there is a lot of computational overkill in fitting a GP, when
a linear model (LM) will fit just as well.  LMs, which can be seen as
a limiting case of the GP, also avoid numerous issues to do with
numerical instability.  It may therefore be desirable to be able to
choose adaptively between a LM and a GP. The goal of this paper is to
link GPs with standard linear models.  One major benefit is the
retention of modeling flexibility while greatly improving the
computational situation.  We can make further gains by combining this
union with treed GPs \citep{gra:lee:2008}, and we demonstrate results
for both treed GPs as well as for stationary GP models.

The remainder of the paper is organized as follows.
Section~\ref{sec:back} reviews the Gaussian process (GP) and the treed
GP.  Section~\ref{sec:llm} introduces the concept of the limiting
linear model (LLM) of a GP.  Therein we argue that, without further
intervention, none of the limiting parameterizations, which are at the
extremes of the parameter space, lead to a feasible model selection
prior.  However, a more thorough exploratory analysis reveals that
there is a broad continuum of GP parameterizations that behave like
the LM and, importantly, tend to have high posterior support when the
data is indeed linear.  It is these parameterizations which we use to
motivate our prior, given in Section~\ref{sec:model}, in order to seek
out a more parsimonious model.  We propose a latent variable
formulation that leads to a prior over a family of semi-parametric
models lying between the GP and its LLM that can be used to
investigate the nature of the influence (in terms of linear versus
non-linear) of predictors on the response.  Section~\ref{sec:results}
covers some details pertaining to efficient implementation, and gives
the results of extensive experimentation on real and synthetic data,
as well as a comparison of our methods to other modern approaches to
regression.  Section~\ref{sec:conclude} concludes with a brief
discussion.

\section{Gaussian Processes and Treed Gaussian Processes}
\label{sec:back}

Consider the following Bayesian hierarchical model for a GP with a
linear mean for $n$
inputs $\mb{X}$ of dimension $m_X$, and $n$ responses $\mb{y}$:
\begin{align}
\mb{y} | \bm{\beta}, \sigma^2, \mb{K} 
        & \sim N_n(\mb{\mb{F}} \bm{\beta}, \sigma^2 \mb{K}) &
        \sigma^2 & \sim IG(\alpha_\sigma/2, q_\sigma/2) \nonumber \\
\bm{\beta} | \sigma^2, \tau^2, \mb{W} 
&\sim N_m(\bm{\beta}_0, \sigma^2 \tau^2 \mb{W}) &
        \tau^2 &\sim IG(\alpha_\tau/2, q_\tau/2) \label{eq:model} \\
\bm{\beta}_0 &\sim N_m(\bm{\mu}, \mb{B}) &
 \mb{W}^{-1} &\sim W((\rho\mb{V})^{-1}, \rho) \nonumber
\end{align}
where $\mb{F} = (\mb{1}, \mb{X})$ has $m=m_X+1$ columns.  $N$, $IG$
and $W$ are the Normal, Inverse-Gamma and Wishart distributions,
respectively.  Constants $\bm{\mu}, \mb{B}, \mb{V}, \rho,
\alpha_\sigma, q_\sigma, \alpha_\tau, q_\tau$ are treated as known.
The matrix $\mb{K}$ is constructed from a function $K(\cdot,\cdot)$ of
the form $K(\mb{x}_j, \mb{x}_k) = K^*(\mb{x}_j, \mb{x}_k) + {g}
\delta_{j,k}$ where $\delta_{\cdot,\cdot}$ is the Kronecker delta
function, $g$ is called the {\em nugget} parameter and is included in
order to interject measurement error (or random noise) into the
stochastic process, and $K^*$ is a true correlation which we take to
be from the separable power family with $p_i = 2$:
\begin{equation} K^*(\mb{x}_j, \mb{x}_k|\mb{d}) = \exp\left\{ -
    \sum_{i=1}^{m_X} \frac{(x_{ij} - x_{ik})^{p_i}}{d_i}\right\}.
\label{e:cor_d} 
\end{equation} 
Generalizations are straightforward, e.g., see Section
\ref{sec:conclude}.  The specification of priors for $K$, $K^*$, and
their parameters $\mb{d}$ and $g$ will be deferred until later, as
their construction will be a central part of this paper.  The
separable power family allows some input variables to be modeled as
more highly correlated than others.  The isotropic power family is a
special case (when $d = d_i$ and $p = p_i$, for $i=1,\dots, m_{X}$).

Posterior inference and estimation is straightforward using the
Metropolis--Hastings (MH) and Gibbs sampling algorithms
\citep{gra:lee:2008}.  It can be shown that the regression
coefficients have full conditionals $\bm{\beta} | \mbox{rest} \sim
N_m(\tilde{\bm{\beta}}, \sigma^2 \mb{V}_{\tilde{\beta}})$, and
$\bm{\beta}_0 | \mbox{rest} \sim N_m(\tilde{\bm{\beta}}_0,
\bm{V}_{\tilde{\beta}_0})$, where
\begin{align} 
\mb{V}_{\tilde{\beta}} &= \label{eq:betavar} 
(\mb{F}^\top\mb{K}^{-1}\mb{F} \!+\! \mb{W}^{-1}/\tau^2)^{-1}, &
  \tilde{\bm{\beta}} &= \mb{V}_{\tilde{\beta}}
  (\mb{F}^\top\mb{K}^{-1} \mb{y} \!+\!
  \mb{W}^{-1}\bm{\beta}_0/\tau^2),\\ 
\mb{V}_{\tilde{\beta}_0} &= \left(\mb{B}^{-1}\! \!+\! \mb{W}^{-1}\!
   \textstyle \sum_{\nu=1}^R (\sigma\tau)^{-2}\right)^{-1}\!\!\!, &
\tilde{\bm{\beta}}_0 &=
\mb{V}_{\tilde{\beta}_0} \left(\mb{B}^{-1}\!\mu \!+\! \mb{W}^{-1}\!
\textstyle \sum_{\nu=1}^R
\bm{\beta} (\sigma\tau)^{-2}\right). \nonumber
\end{align} 
%The linear variance parameter has full conditional
%\begin{equation}
%\tau^2 | \mbox{rest} \sim IG((\alpha_\tau + m)/2, (q_\tau +
%b)/2), 
%\mbox{\hspace{0.2cm} where \hspace{0.1cm}} b = (\bm{\beta} -
%\bm{\beta}_0)^\top \mb{W}^{-1} (\bm{\beta} - \bm{\beta}_0)/\sigma^2.
%\nonumber
%\end{equation} 
%The linear model covariance matrix $\mb{W}$ follows an inverse-Wishart:
%\begin{equation} 
%\mb{W}^{-1}|\mbox{rest} \sim
%W_m\left((\rho \mb{V}\mb + \mb{V}_{\hat{W}})^{-1}, \rho+R\right), 
%\mbox{\hspace{0.2cm} where \hspace{0.1cm}} \mb{V}_{\hat{W}} = \sum_{\nu=1}^R
%\frac{1}{(\sigma\tau)^2} (\bm{\beta} - \bm{\beta}_0)
%(\bm{\beta} - \bm{\beta}_0)^\top.  \nonumber
%\end{equation}
%The above full conditionals allow Gibbs sampling.
Analytically integrating out $\bm{\beta}$ and $\sigma^2$ gives
a marginal posterior for $\mb{K}$ \citep{berg:deol:sans:2001} that
can be used to obtain efficient MH draws.  
\begin{align} 
p(\mb{K} |\mb{y},\bm{\beta}_0, \mb{W}, \tau^2) = %\nonumber \\ 
& \left(\frac{|\mb{V}_{\tilde{\beta}}|(2\pi)^{-n}}{ 
        |\mb{K}||\mb{W}|\tau^{2m}}\right)^{\frac{1}{2}}
        \frac{\left(\frac{q_\sigma}{2}\right)^{\alpha_\sigma/2}
        \Gamma\left[\frac{1}{2}(\alpha_\sigma + n)\right]}
        {\left[\frac{1}{2}(q_\sigma+\psi)\right]^{(\alpha_\sigma+n)/2}
        \Gamma\left[\frac{\alpha_\sigma}{2}\right]} p(\mb{K}), 
        \label{e:marginp}
\end{align} 
\vspace*{-0.75cm}
\[
%\begin{equation}
\mbox{where \hspace{1cm}} \psi = \mb{y}^\top \mb{K}^{-1} \mb{y} +
\bm{\beta}_0^\top \mb{W}^{-1} \bm{\beta}_0/\tau^2 - 
\tilde{\bm{\beta}}^\top
\mb{V}_{\tilde{\beta}}^{-1} \tilde{\bm{\beta}}.
%\label{eq:phi}
%\end{equation}
\]
%Any hyperparameters to $K(\cdot, \cdot)$, e.g., parameters to priors
%for $\{d,g\}$ of the isotropic power family, would also require MH
%draws.  The conditional distribution of $\sigma^2$ with $\bm{\beta}$
%integrated out allows Gibbs sampling:
%\begin{equation} 
%  \sigma^2 | \mb{y}, d, g, \bm{\beta}_0, \mb{W} \sim 
%  IG((\alpha_\sigma + n)/2, (q_\sigma + \psi)/2).
%  \label{e:s2marginp} 
%\end{equation} 
%
The predicted value of $y$ at $\mb{x}$ is normally distributed with
mean and variance
\begin{align} 
\hat{y}(\mb{x}) &\!=\! \mb{f}^\top(\mb{x}) \tilde{\bm{\beta}} \!+\!
        \mb{k}(\mb{x})^\top \mb{K}^{-1}(\mb{y} \!-\!
        \mb{F}\tilde{\bm{\beta}}), \label{eq:pred}& % \nonumber \\
\hat{\sigma}(\mb{x})^2 \!&=\! \sigma^2 [\kappa(\mb{x},
\mb{x}) \!-\! \mb{q}^\top(\mb{x})\mb{C}^{-1} \mb{q}(\mb{x})],
\end{align} 
where $\tilde{\bm{\beta}}$ is the posterior mean estimate of
$\bm{\beta}$, $\mb{C}^{-1} = (\mb{K} \!+\! \tau^2\mb{F}
\mb{W}\mb{F}^\top)^{-1}$, $\mb{q}(\mb{x}) = \mb{k}(\mb{x}) \!+\!
\tau^2\mb{F} \mb{W} \mb{f}(\mb{x})$, $\kappa(\mb{x},\mb{y}) =
K(\mb{x},\mb{y}) \!+\! \tau^2\mb{f}^\top(\mb{x}) \mb{W} \mb{f}(\mb{y})$,
$\mb{f}^\top(\mb{x}) = (1, \mb{x}^\top)$, and $\mb{k}(\mb{x})$ is a
$n-$vector with $\mb{k}(\mb{x})_j= K(\mb{x}, \mb{x}_j)$, for all
$\mb{x}_j \in \mb{X}$, the training data.

A treed GP \citep{gra:lee:2008} is a generalization of the CART
(Classification and Regression Tree) model \citep{chip:geor:mccu:1998}
that uses GPs at the leaves of the tree in place of the usual constant
values or the linear regressions of \cite{chip:geor:mccu:2002}.  The
Bayesian interpretation requires a prior be placed on the 
tree and GP parameterizations.  Sampling commences with Reversible
Jump (RJ) MCMC which allows for a simultaneous fit of the tree and the
GPs at its leaves.  The predictive surface can be discontinuous across
the partition boundaries of a particular tree $\mathcal{T}$.  However,
in the aggregate of samples collected from the joint posterior
distribution of $\{\mathcal{T}, \bm{\theta}\}$, the mean tends to
smooth out near likely partition boundaries as the RJ--MCMC integrates
over trees and GPs according to the posterior distribution
\citep{gra:lee:2008}.  %Uncertainty in the posterior for $\mathcal{T}$
%translates into higher posterior predictive uncertainty near region
%boundaries.

The treed GP approach yields an extremely fast implementation of
nonstationary GPs, providing a divide-and-conquer approach to spatial
modeling.  Software implementing the treed GP model and all of its
special cases (e.g., stationary GP, CART \& the treed linear model,
linear model, etc.), including the extensions discussed in this paper,
is available as an {\sf R} package~\citep{cran:R}, and can be obtained
from CRAN:
\begin{center}
\verb!http://www.cran.r-project.org/web/packages/tgp/index.html!.
\end{center}
The package implements a family of default prior specifications for
the known constants in Eq.~(\ref{eq:model}), and those described in
the following sections, which are used throughout unless otherwise
noted.  For more details see the {\tt tgp} documentation
\citep{package:tgp} and tutorial \citep{Gramacy:2007:JSSOBK:v19i09}.

\section{Limiting Linear Models}
\label{sec:llm}

A special limiting case of the GP model is the standard linear model
(LM).  Replacing the top (likelihood) line in the hierarchical model
given in (\ref{eq:model})
\begin{align*} 
  \mb{y} | \bm{\beta}, \sigma^2, \mb{K} &\sim N_n(\mb{\mb{F}}
  \bm{\beta}, \sigma^2 \mb{K}) && \mbox{with}& \mb{y} | \bm{\beta},
  \sigma^2 &\sim N_n(\mb{\mb{F}} \bm{\beta}, \sigma^2 \mb{I}),
\end{align*} where $\mb{I}$ is the $n \times n$ identity matrix, gives
a parameterization of a LM.  From a phenomenological perspective, GP
regression is more flexible than standard linear regression in that it
can capture nonlinearities in the interaction between covariates
($\mb{x}$) and responses ($y$).  From a modeling perspective, the GP
can be more than just overkill for linear data.  Parsimony and
over-fitting considerations are just the tip of the iceberg.  It is
also unnecessarily computationally expensive, as well as numerically
unstable.  Specifically, it requires the inversion of a large
covariance matrix---an operation whose computing cost grows with the
cube of the sample size, $n$.  Moreover, large finite $\mb{d}$
parameters can be problematic from a numerical perspective because,
unless $g$ is also large, the resulting covariance matrix can be
numerically singular when the off-diagonal elements of $\mb{K}$ are
nearly one.

It is common practice to scale the inputs ($\mb{x}$) either to lie in
the unit cube, or to have a mean of zero and a range of one.  Scaled
data and mostly linear predictive surfaces can result in almost
singular covariance matrices even when the range parameter is
relatively small ($2 < d_i \ll \infty$).  So for some parameterizations,
the GP is operationally equivalent to the limiting linear model (LLM),
but comes with none of its benefits (e.g. speed and stability).  As
this paper demonstrates, exploiting and/or manipulating such
equivalence can be of great practical benefit.  As Bayesians, this
means constructing a prior distribution on $\mb{K}$ that makes it
clear in which situations each model is preferred (i.e., when should
$\mb{K} \rightarrow c\mb{I}$?).  Our key idea is to specify a prior on
a ``jumping'' criterion between the GP and its LLM by taking advantage
of a latent variable formulation, thus setting up a Bayesian model
selection/averaging framework.

Theoretically, there are only two parameterizations of a GP
correlation structure ($K$) which encode the LLM.  Though they are
indeed well--known, without intervention they are quite unhelpful from
the perspective of {\em practical} estimation and inference.  The
first one is when the range parameter ($\mb{d}$) is set to zero.  In this
case $\mb{K} = (1+g)\mb{I}$, and the result is clearly a linear model.
The other parameterization may be less obvious.

\citet[Section 3.2.1]{cressie:1991} analyzes the ``effect of variogram
parameters on kriging'' paying special attention to the nugget ($g$)
and its interaction with the range parameter.  He remarks that the
larger the nugget the more the kriging interpolator smoothes and in
the limit predicts with the linear mean.  Perhaps more relevant to the
forthcoming discussion is his later remarks on the interplay between
the range and nugget parameters in determining the {\em kriging
  neighborhood}.  Specifically, a large nugget coupled with a large
range drives the interpolator towards the linear mean.  This is
refreshing since constructing a prior for the LLM by exploiting the
former GP parameterization (range $\mb{d}\rightarrow \mb{0}$) is
difficult, and for the latter (nugget $g\rightarrow\infty$) near
impossible.  We regard these parameterizations, which are situated at
the extremes of the parameter space, as a dead--end as far as serving
as the basis for a model--selection prior.  Fortunately, Cressie's
thoughts on the kriging neighborhood reveal that an (essentially)
linear model may be attainable with nonzero $\mb{d}$ and finite $g$.

\subsection{Exploratory analysis}
\label{sec:gpllm:explore}

Here we shall conduct an exploratory analysis to study the kriging
neighborhood and look for a platform from which to ``jump'' to the
LLM.  The analysis will focus on studying likelihoods and posteriors
for GPs fit to data generated from the linear model
\begin{equation} 
y_i = 1 + 2x_i + \epsilon_, \;\;\;\;\; \mbox{where} \;\;\;
\epsilon_i \stackrel{\mbox{\tiny iid}}{\sim} N(0,1) 
\label{eq:linear:sim}
\end{equation} 
using $n=10$ evenly spaced $x$-values in the range $[0,1]$.

\subsubsection{GP likelihoods on linear data}
\label{sec:gpllm:gpliklin}

\begin{figure}[ht!] 
\begin{center} 
\begin{tabular}{lr} 
\includegraphics[scale=0.23,angle=-90,trim=0 35 0 0]{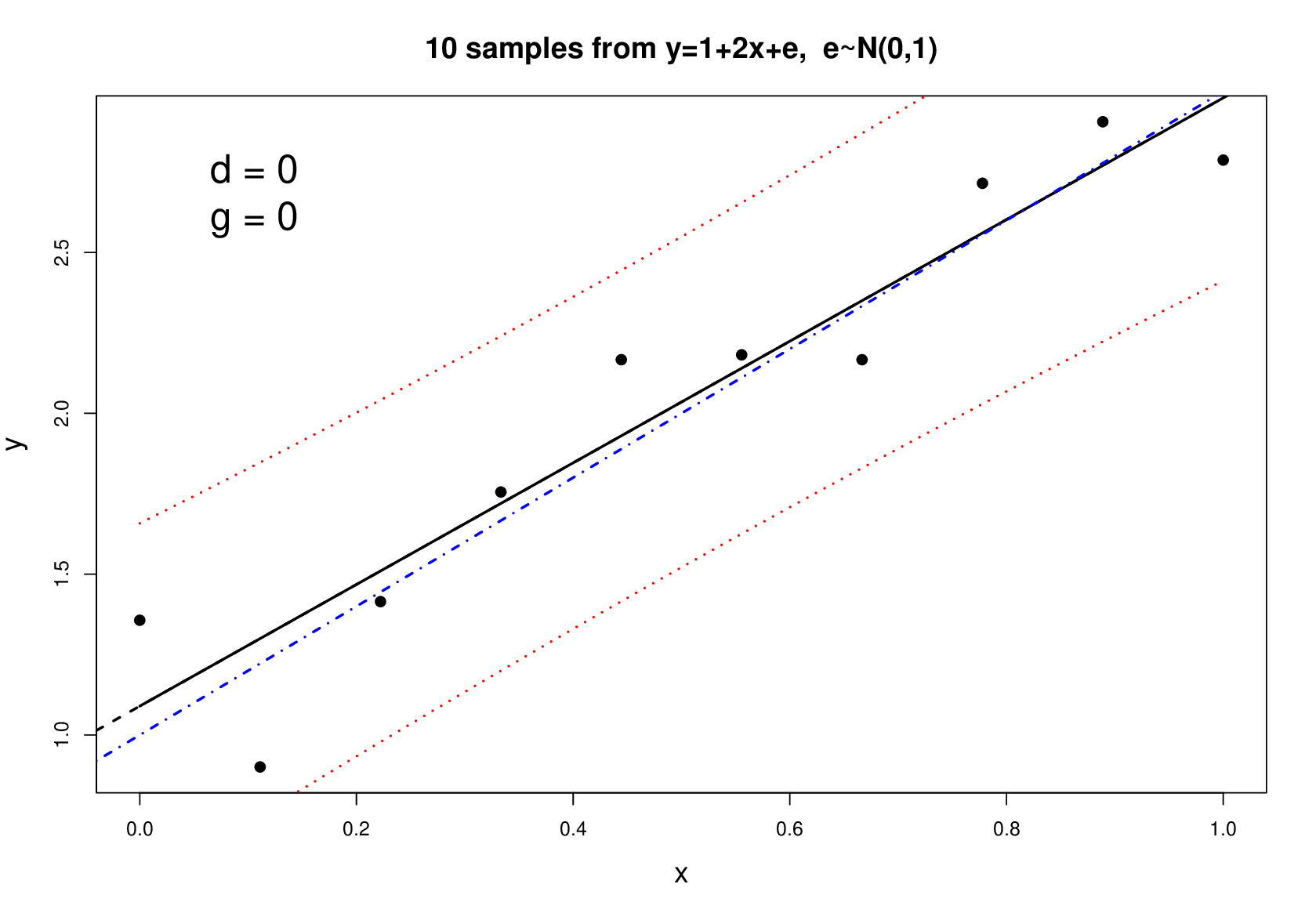} & 
\includegraphics[scale=0.23,angle=-90,trim=0 35 0 0]{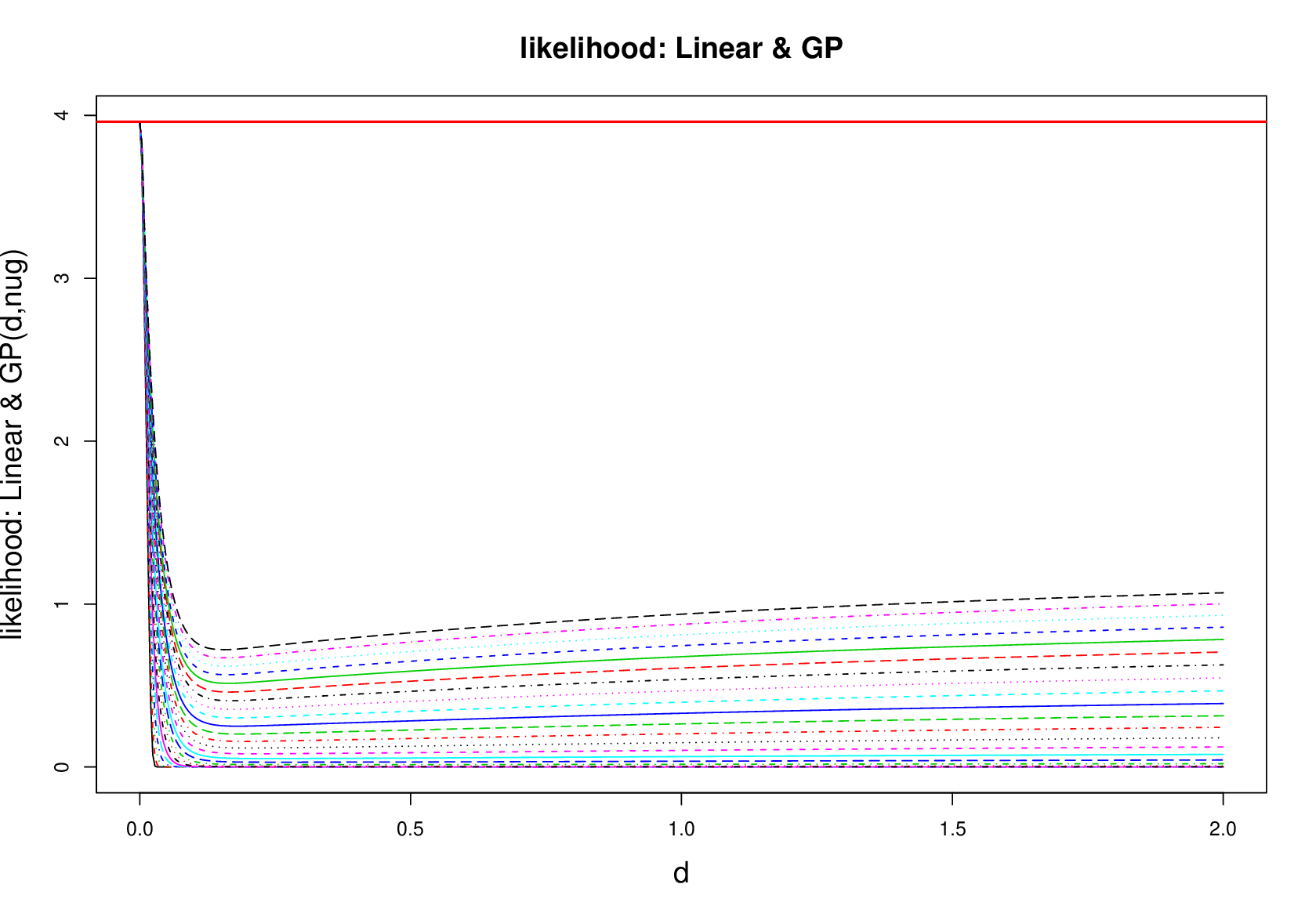} \\
\includegraphics[scale=0.23,angle=-90,trim=0 35 0 0]{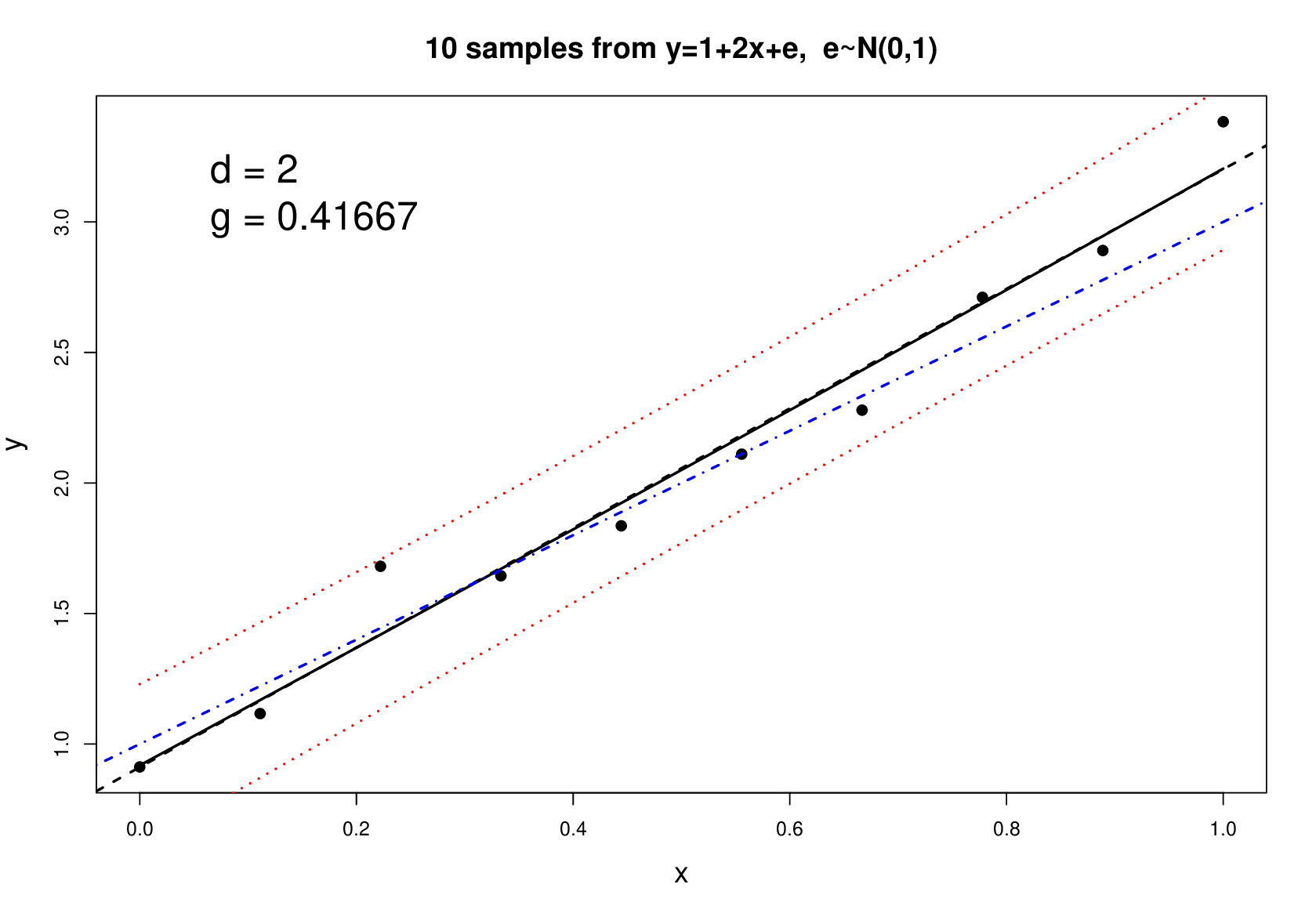} &
\includegraphics[scale=0.23,angle=-90,trim=0 35 0 0]{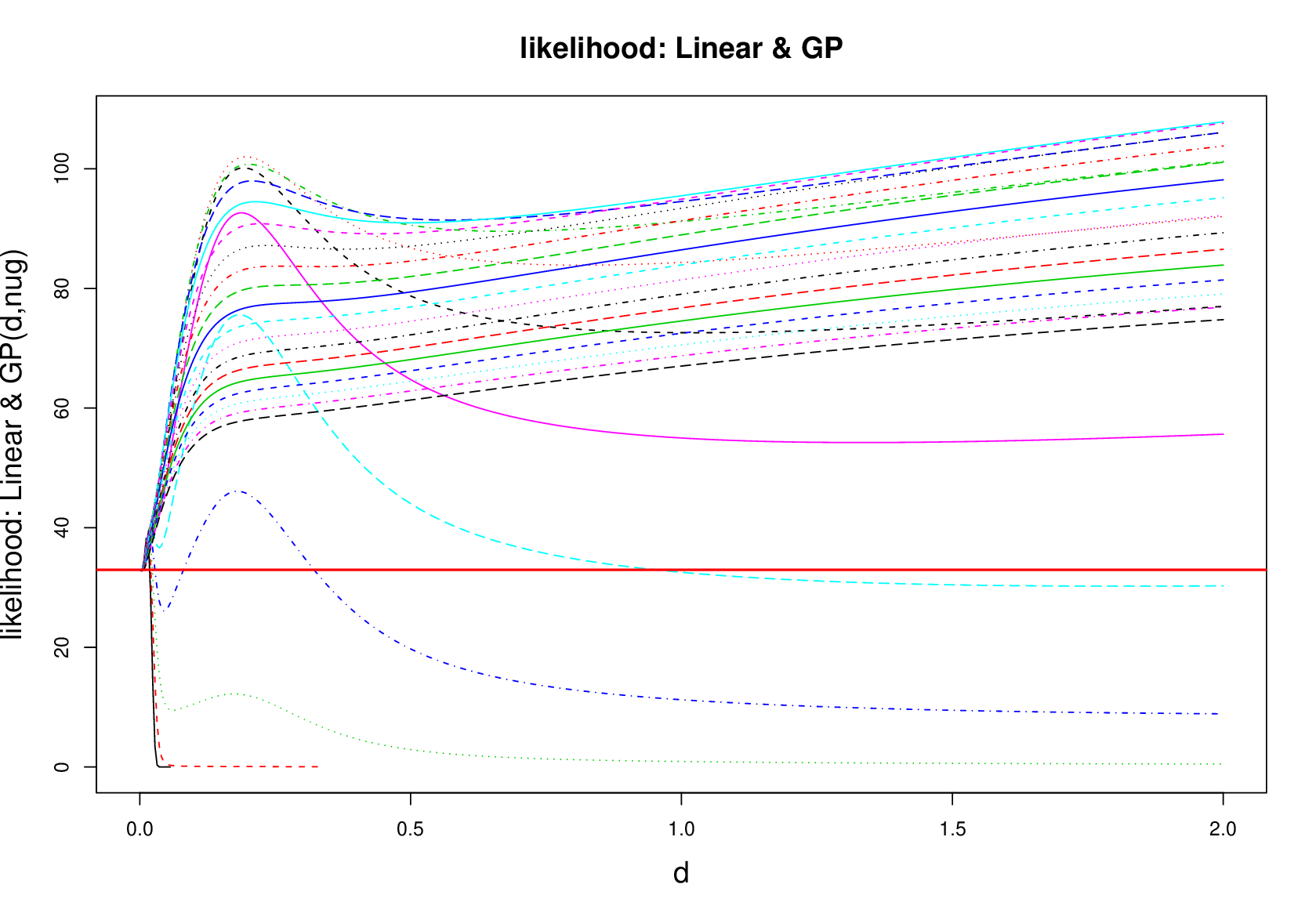}
\end{tabular} 
\caption[ML GP fits of two samples from a linear model] {Two
  simulations ({\em rows}) from $y_i=1+2x_i+\epsilon_i, \;
  \epsilon_i\sim N(0,1)$.  {\em Left} column shows GP fit (solid) with
  95\% error bars (dotted), maximum likelihood $\hat{\bm{\beta}}$
  (dashed), and generating linear model ($\bm{\beta} = (1,2)$)
  (dot-dashed).  {\em Right} column shows GP$(d,g)$ likelihood
  surfaces, with each curve representing a different value of the
  nugget $g$.  The (maximum) likelihood $(\hat{\bm{\beta}})$ of the
  linear model is indicated by the solid horizontal line.  }
\label{f:gpvlin}
\end{center} 
\end{figure}

Figure~\ref{f:gpvlin} shows two interesting samples from
(\ref{eq:linear:sim}).  Also plotted are the generating line
(dot-dashed), the maximum likelihood (ML) linear model
($\hat{\bm{\beta}}$) line (dashed), the predictive mean surface of the
ML GP, maximized over range $d$, i.e., the one--vector $\mb{d}$,
nugget $g$, and $[\sigma^2|d,g]$ (solid), and its 95\% error bars
(dotted).  The ML values of $d$ and $g$ are also indicated in each
plot.  The GP likelihoods were evaluated for ML estimates of the
regression coefficients $\hat{\bm{\beta}}$.  Conditioning on $g$ and
$d$, the ML variance was computed by solving
\[
0 \equiv \frac{d}{d\sigma^2} 
        \log N_n(\mb{y} | \mb{F} \hat{\bm{\beta}}, \sigma^2 \mb{K}) \nonumber 
= -\frac{n}{\sigma^2} + 
        \frac{(\mb{y} - \mb{F}\hat{\bm{\beta}})^\top \mb{K}^{-1}
        (\mb{y} - \mb{F}\hat{\bm{\beta}})}{(\sigma^2)^2}.
\]
This gave an MLE with the form $\hat{\sigma}^2 = (\mb{y} -
\mb{F}\hat{\bm{\beta}})^\top \mb{K}^{-1} (\mb{y} -
\mb{F}\hat{\bm{\beta}})/n$.  For the linear model the likelihood was
evaluated as $P(\mb{y}) = N_{10} (\mb{F}\hat{\bm{\beta}},
\hat{\sigma}^2 \mb{I})$, and for the GP as $ P(\mb{y}|d,g) = N_{10}
\left[\mb{F}\hat{\bm{\beta}}, \hat{\sigma}^2 \mb{K}_{\{d,g\}}\right]$,
where $\mb{K}_{\{d,g\}}$ is the covariance matrix generated using
$K(\cdot, \cdot) = K^*(\cdot, \cdot| d) + g\delta_{\cdot,\cdot}$ for $
K^*(\cdot, \cdot| d)$ from the power family with range parameter $d$.

Both samples and fits plotted in Figure~\ref{f:gpvlin} have linear
looking predictive surfaces, but only for the one in the {\em top} row
does the linear model have the maximum likelihood.  Though the
predictive surface in the {\em bottom-left} panel could be mistaken as
``linear'', it was indeed generated from a GP with large range
parameter ($d = 2$) and modest nugget setting ($g$) as this
parameterization had higher likelihood than the linear model.  The
{\em right} column of Figure~\ref{f:gpvlin} shows likelihood surfaces
corresponding to the samples in the {\em left} column.  Each curve
corresponds to a different setting of the nugget, $g$.  Also shown is
the likelihood value of the MLE $\hat{\bm{\beta}}$ of the linear model
(solid horizontal line).  The likelihood surfaces for each sample look
drastically different.  In the {\em top} sample the LLM ($d=0$)
uniformly dominates all other GP parameterizations.  Contrast this
with the likelihood of the second sample.  There, the resulting
predictive surface looks linear, but the likelihood of the LLM is
comparatively low.

\begin{figure}[ht!] 
\begin{center}
\begin{tabular}{lcr}
\includegraphics[scale=0.28,angle=-90]{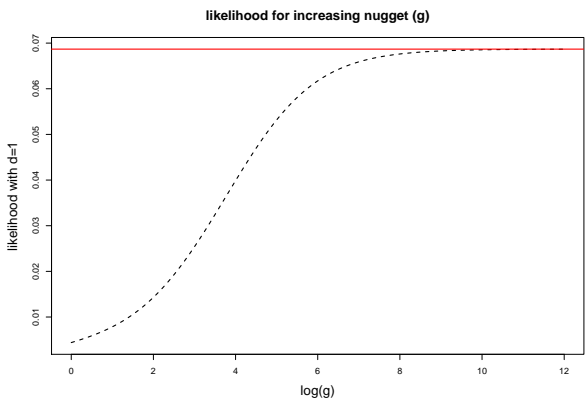}
\end{tabular} 
\caption[GP likelihoods on linear data as nugget gets large]
{Likelihoods as the nugget gets 
large for single sample of size $n=100$ from Eq.~(\ref{eq:linear:sim}).  
The $x$-axis is $(\log g)$, the range is fixed at $d=1$;
the likelihood of the LLM ($d=0$) is shown for comparison.
\label{f:bignug}}
\end{center} 
\vspace{-0.8cm}
\end{figure}

Figure \ref{f:bignug} illustrates the other LLM parameterization by
showing how, as the nugget $g$ increases, the likelihood of the GP
approaches that of the linear model for a sample of size $n=100$ from
(\ref{eq:linear:sim}) with $d$ fixed to one.  Observe that the nugget
must be quite large relative to the actual variability in the data
before the likelihoods of the GP and LLM become comparable.

\begin{figure}[ht!] 
\begin{center}
\begin{tabular}{lr}
\includegraphics[scale=0.22,angle=-90,trim=0 35 0 0]{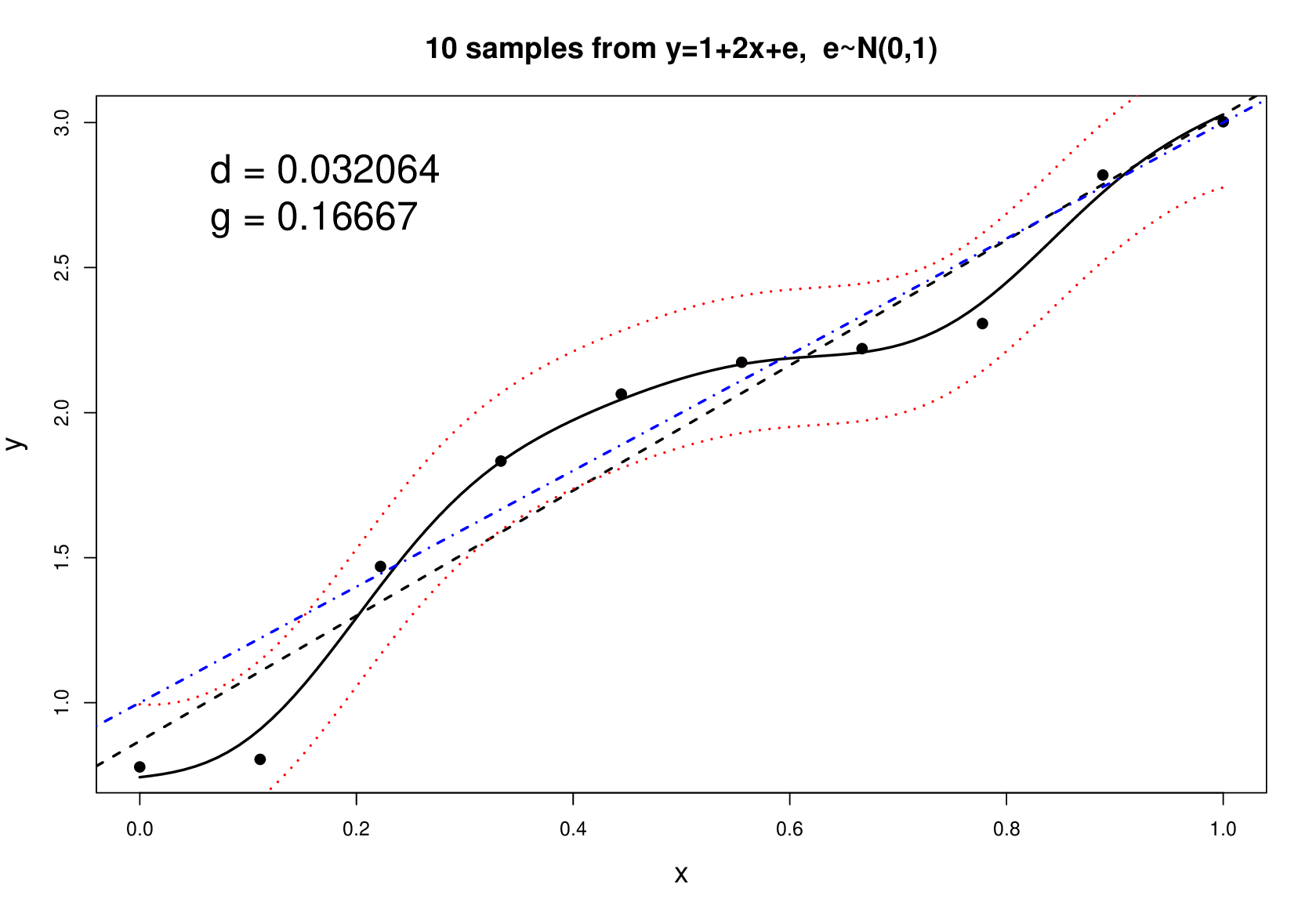} &
\includegraphics[scale=0.22,angle=-90,trim=0 35 0 0]{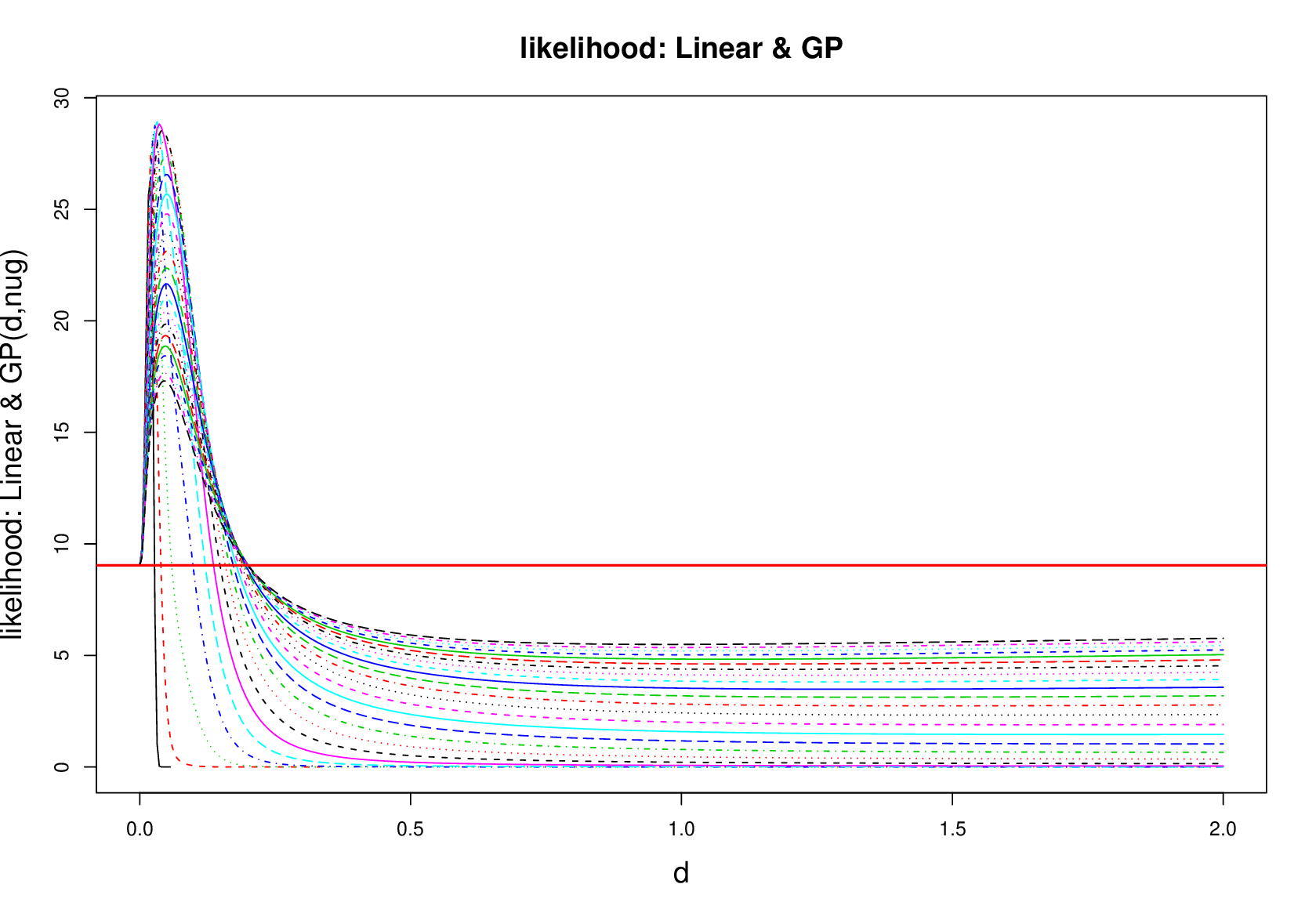} \\
\includegraphics[scale=0.22,angle=-90,trim=0 35 0 0]{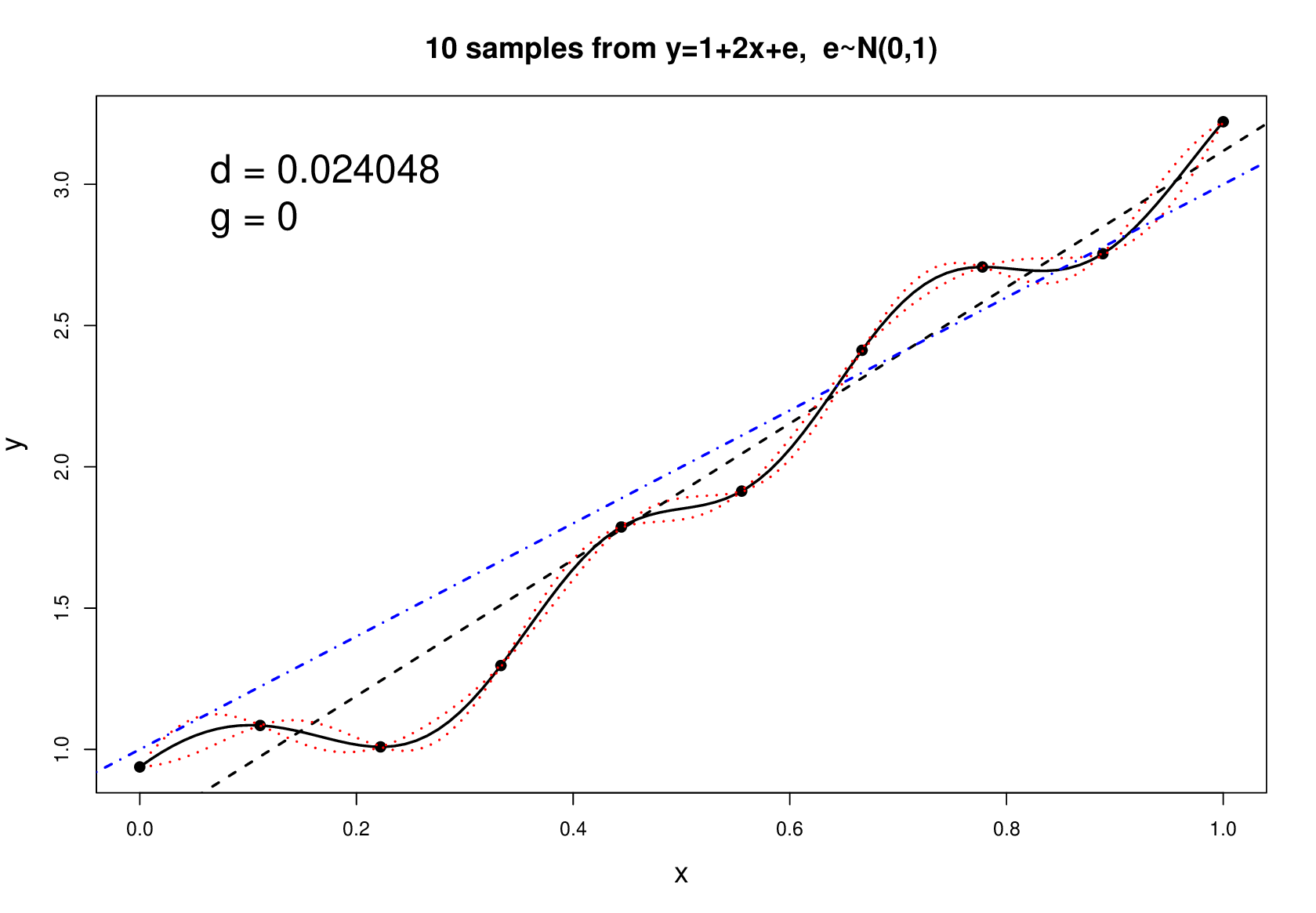} &
\includegraphics[scale=0.22,angle=-90,trim=0 35 0 0]{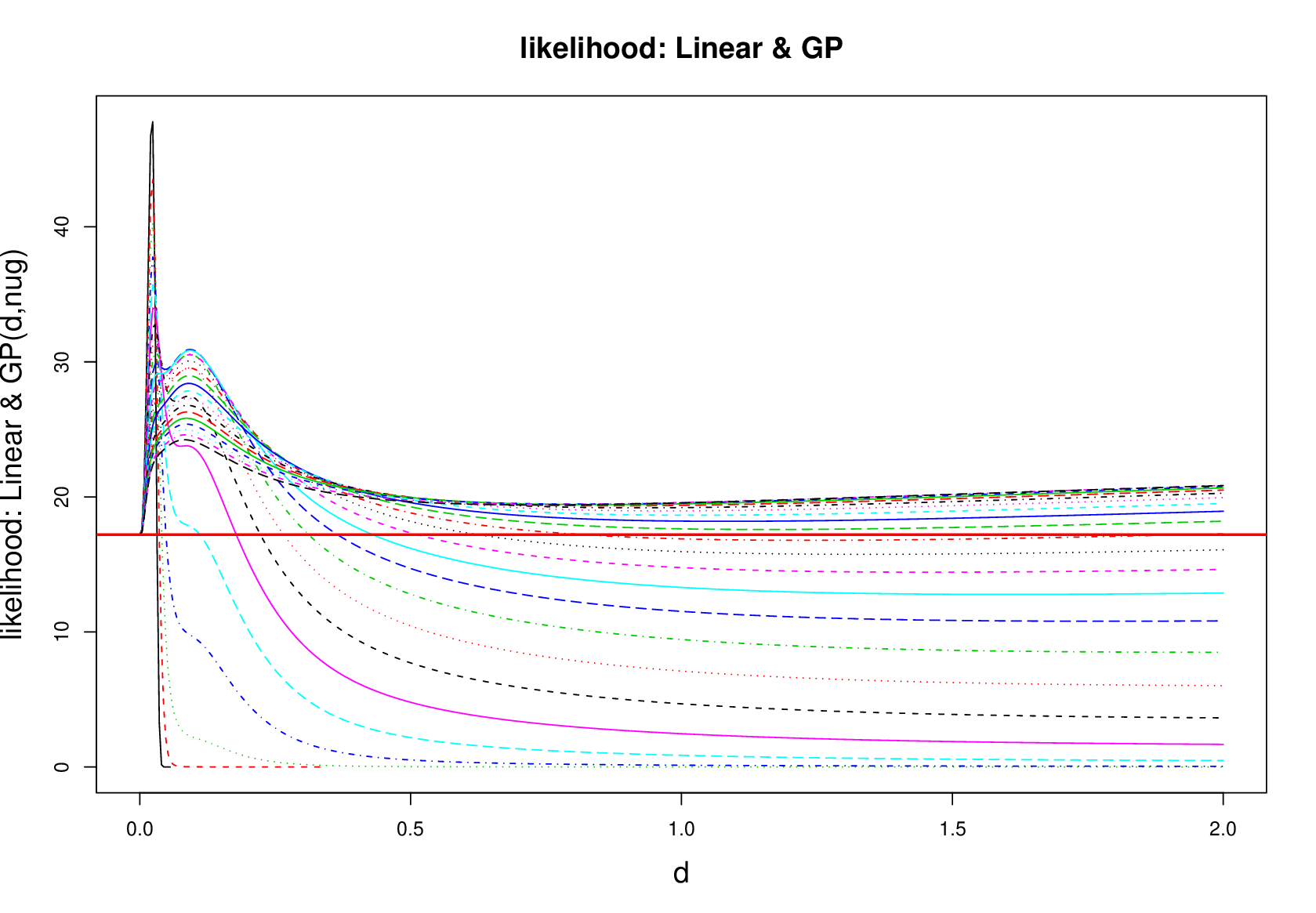}
\end{tabular} \caption[Wiggly ML GP fits to linear data]
{GP$(d,g)$ fits ({\em left}) and
likelihood surfaces ({\em right}) for two of
samples of the LM (\ref{eq:linear:sim}). 
\label{f:wavy}} 
\end{center} 
\end{figure}

Most simulations from (\ref{eq:linear:sim}) gave predictive surfaces
like the {\em upper left} of Figure \ref{f:gpvlin} and corresponding
likelihoods like the {\em upper-right}.  But this is not always the
case.  Occasionally a simulation would give high likelihood to GP
parameterizations if the sample was slowly waving.  This is not
uncommon for small sample sizes such as $n=10$.  For example, consider
those shown in Figure \ref{f:wavy}.  Waviness becomes less likely as
the sample size $n$ grows.

\begin{figure}[ht!] 
\begin{center} 
\begin{minipage}{4cm} 
\begin{center}
\includegraphics[scale=0.28,angle=-90]{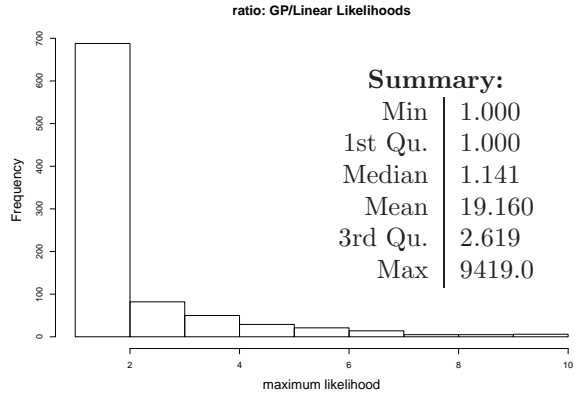} 
\end{center}
\end{minipage} 
\begin{minipage}{5.5cm} 
\footnotesize
\begin{tabular}{r|l}
\multicolumn{2}{c}{\bf Summary:} \\ Min & 1.000 \\ 1st Qu. & 1.000 \\ Median &
1.141 \\ Mean & 19.160 \\ 3rd Qu. & 2.619 \\ Max & 9419.0 
\end{tabular}
\end{minipage} 
\caption[Histogram of GP/LM likelihoods on linear data]
{Histogram of the ratio of the maximum
likelihood GP parameterization over the likelihood of the limiting linear
model with full summary statistics. For visual clarity, the upper tail
of the histogram is not shown.} \label{f:gpvlin:hist}
\end{center} 
\end{figure}

Figure \ref{f:gpvlin:hist} summarizes the ratio of the ML GP
parameterization over the ML LM based on 1000 simulations of ten
evenly spaced random draws from (\ref{eq:linear:sim}).  A likelihood
ratio of one means that the LM was best for a particular sample.  The
90\%-quantile histogram and summary statistics in Figure
\ref{f:gpvlin:hist} show that the GP is seldom much better than the
LM.  For some samples the ratio can be really large ($>9000$) in favor
of the GP, but more than two-thirds of the ratios are close to
one---approximately 1/3 (362) were exactly one but 2/3 (616) had
ratios less than 1.5.  This means that the posterior inference for
borderline linear data is likely to depend heavily the prior
specification of $K(\cdot,\cdot)$.

For some of the smaller nugget values, in particular $g=0$, and larger
range settings $d$, the likelihoods for the GP could not be computed
because the corresponding covariance matrices were numerically
singular, and could not be decomposed.  This illustrates a phenomenon
noted by Neal (1997) \nocite{neal:1997} who advocates that a non-zero
nugget (or {\em jitter}) should be included in the model to increase
numerical stability.  Numerical instabilities may also be avoided by
allowing $p_i < 2$ in (\ref{e:cor_d}), by using the Mat\'{e}rn family
of correlation functions (see Section \ref{sec:conclude}), or by
simply using an LM where appropriate.

\subsubsection{GP posteriors on linear data}
\label{sec:gpllm:gppostlin}

Suppose that rather than examining the multivariate normal likelihoods
of the linear and GP model, using the ML $\hat{\bm{\beta}}$ and
$\hat{\sigma}^2$ values, the marginalized posterior $p(\mb{K} |
\mb{y}, \bm{\beta}_0, \tau^2, \mb{W})$ of Eq.~(\ref{e:marginp}) was
used, which integrates out $\bm{\beta}$ and $\sigma^2$.  Using
(\ref{e:marginp}) requires specification of the prior $p(\mb{K})$,
which for the power family means specifying $p(d,g)$.  Alternatively,
one could consider dropping the $p(d,g)$ term from (\ref{e:marginp})
and look solely at the marginalized likelihood.  However, in light of
the arguments above, there is reason to believe that the prior
specification will carry significant weight.

If it is suspected that the data might be linear, this bias should be
somehow encoded in the prior.  This is a non-trivial task, given the
nature of the GP parameterizations which encode the LLM.  Pushing $d$
towards zero is problematic because small non-zero $d$ causes the
predictive surface to be quite wiggly---certainly far from linear.
Deciding how small the range parameter ($d$) should be before treating
it as zero---as in Stochastic Search Variable Selection (SSVS) of
\citet{geor:mccu:1993}, or Chapter 12 of \cite{gilks:1996}---while
still allowing a GP to fit truly non-linear data is no simple task.
The large nugget approach is also out of the question because putting
increasing prior density on a parameter as it gets large is
impossible.  Rescaling the responses might work, but constructing the
prior would be nontrivial.  Moreover, such an approach would preclude
its use in many applications, particularly for adaptive sampling or
sequential design of experiments when one hopes to learn about the
range of responses, and/or search for extrema.

However, we have seen that for a continuum of large $d$ values (say
$d>0.5$ on the unit interval) the predictive surface is practically
linear.  Consider a mixture of gammas prior for $d$:
\begin{align} 
p(d,g) &= p(d) p(g)
= p(g) [G(d|\alpha=1,\beta=20) + G(d|\alpha=10,\beta=10)]/2.
\label{eq:dprior}
\end{align}  
It gives roughly equal mass to small $d$ (mean $1/20$) representing a
population of GP parameterizations for wavy surfaces, and a separate
population for those which are quite smooth or approximately linear.
Figure \ref{f:boolprior} depicts $p(d)$ via a histogram, ignoring
$p(g)$ which is usually taken to be a simple exponential distribution.
Alternatively, one could encode the prior as $p(d,g) = p(d|g)p(g)$ and
then use a reference prior \citep{berg:deol:sans:2001} for $p(d|g)$.
We chose the more deliberate, independent, specification in order to
encode our prior belief that there are essentially two kinds of
processes: wavy (small $d$) and smooth (large $d$).  Observe that $d =
0$ is ``closer'' to the wavy parameterizations---in fact it lies at
the limit of extreme waviness.  We argue, however, that extreme
smoothness may not only be a more intuitive depiction of linearity,
but that large $d$-values (though further away) serve as a better
platform from which to jump to the LLM.

Evaluation of the marginalized posterior (\ref{e:marginp}) requires
settings for the prior mean coefficients $\bm{\beta}_0$, covariance
$\tau^2 \mb{W}$, and hierarchical specifications $(\alpha_\sigma,
\gamma_\sigma)$ for $\sigma^2$.  For now, these parameter settings are
fixed to those which were known to generate the data.

\begin{figure}[ht!] 
\begin{center}
\includegraphics[scale=0.205,trim=20 50 10 27,angle=-90]{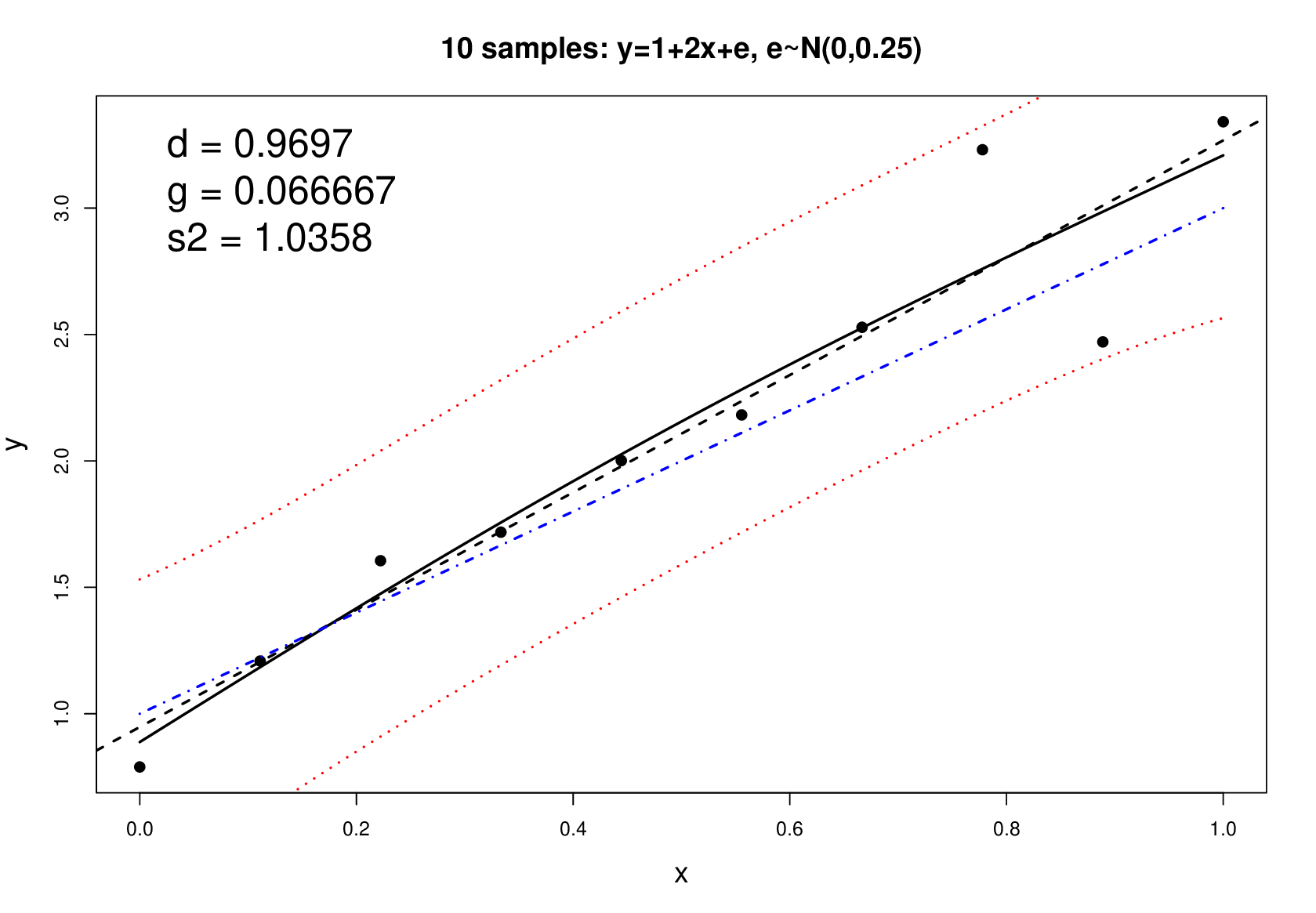}
\includegraphics[scale=0.205,trim=20 20 10 27,clip=TRUE,angle=-90]{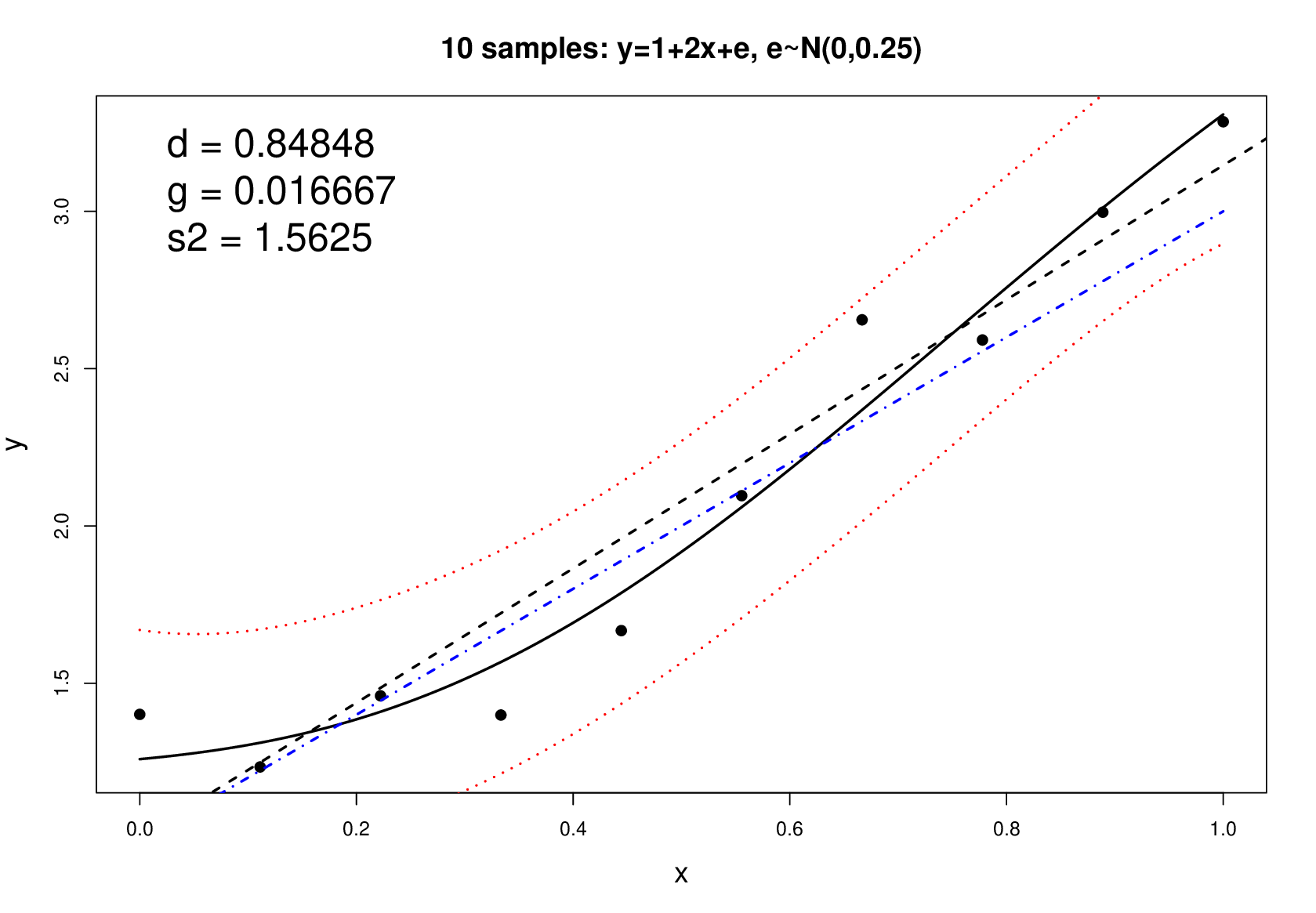}
\includegraphics[scale=0.205,trim=20 20 10 27,clip=TRUE,angle=-90]{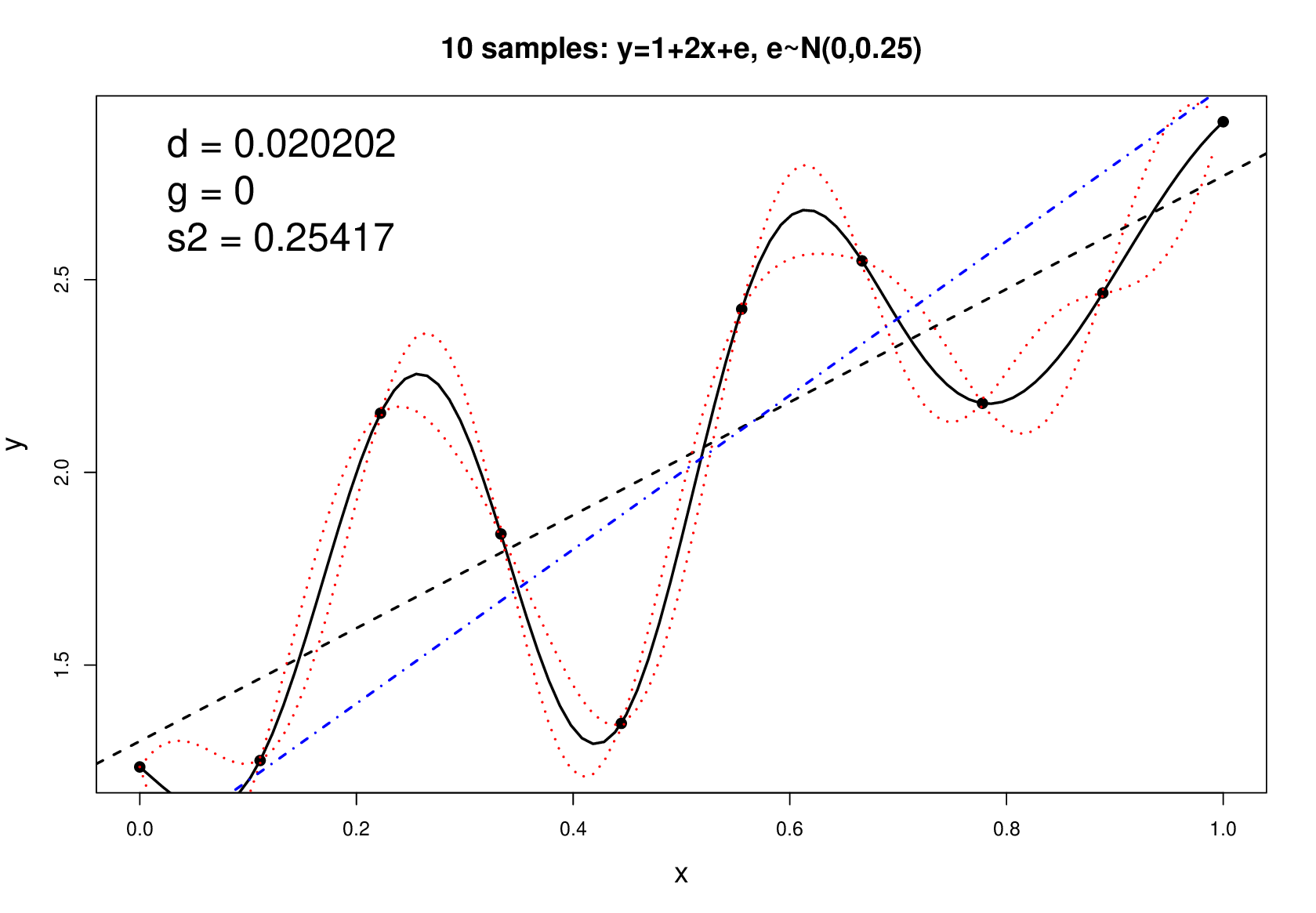}
\includegraphics[scale=0.205,trim=20 50 10 27,angle=-90]{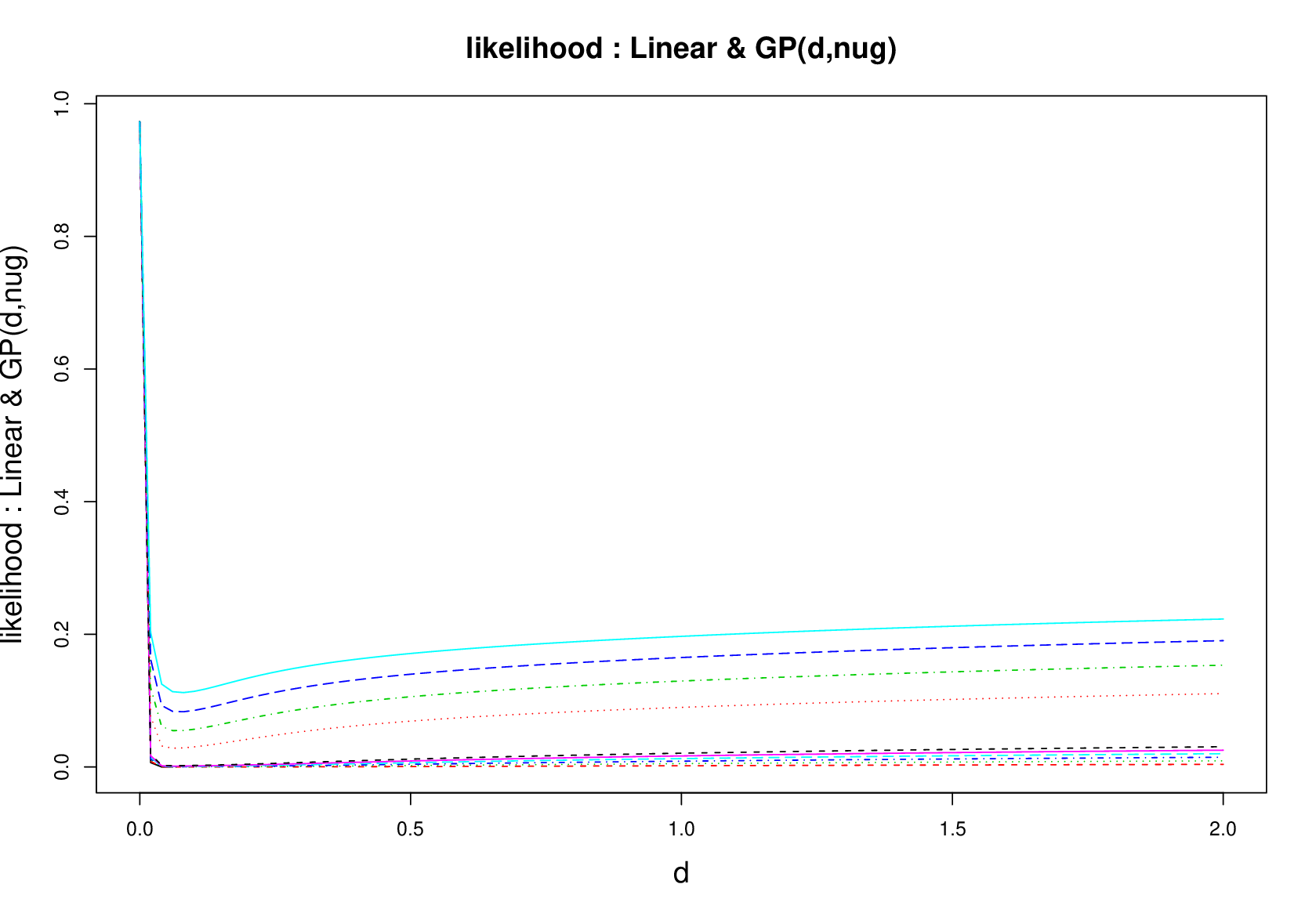}
\includegraphics[scale=0.205,trim=20 20 10 27,clip=TRUE,angle=-90]{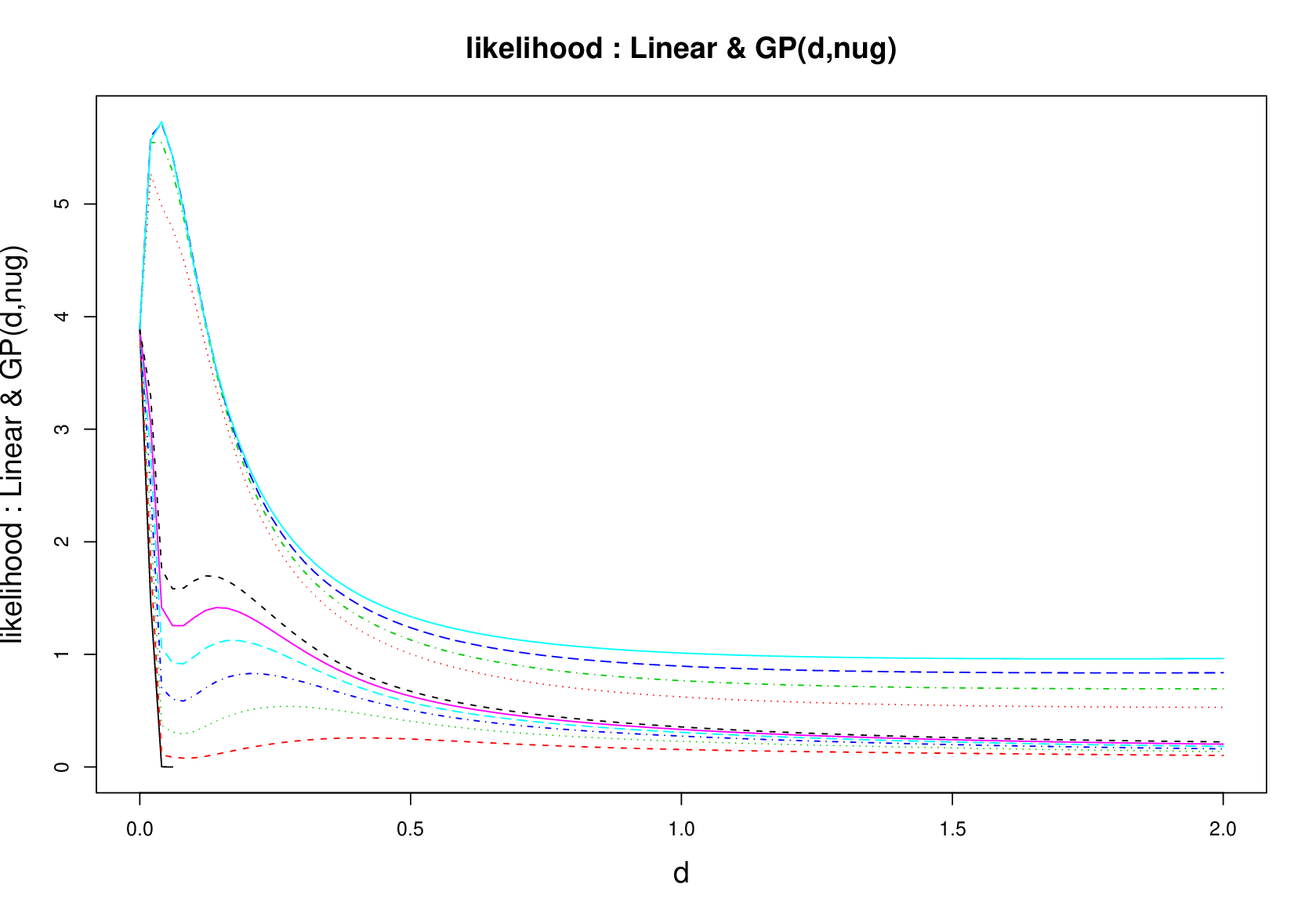}
\includegraphics[scale=0.205,trim=20 20 10 27,clip=TRUE,angle=-90]{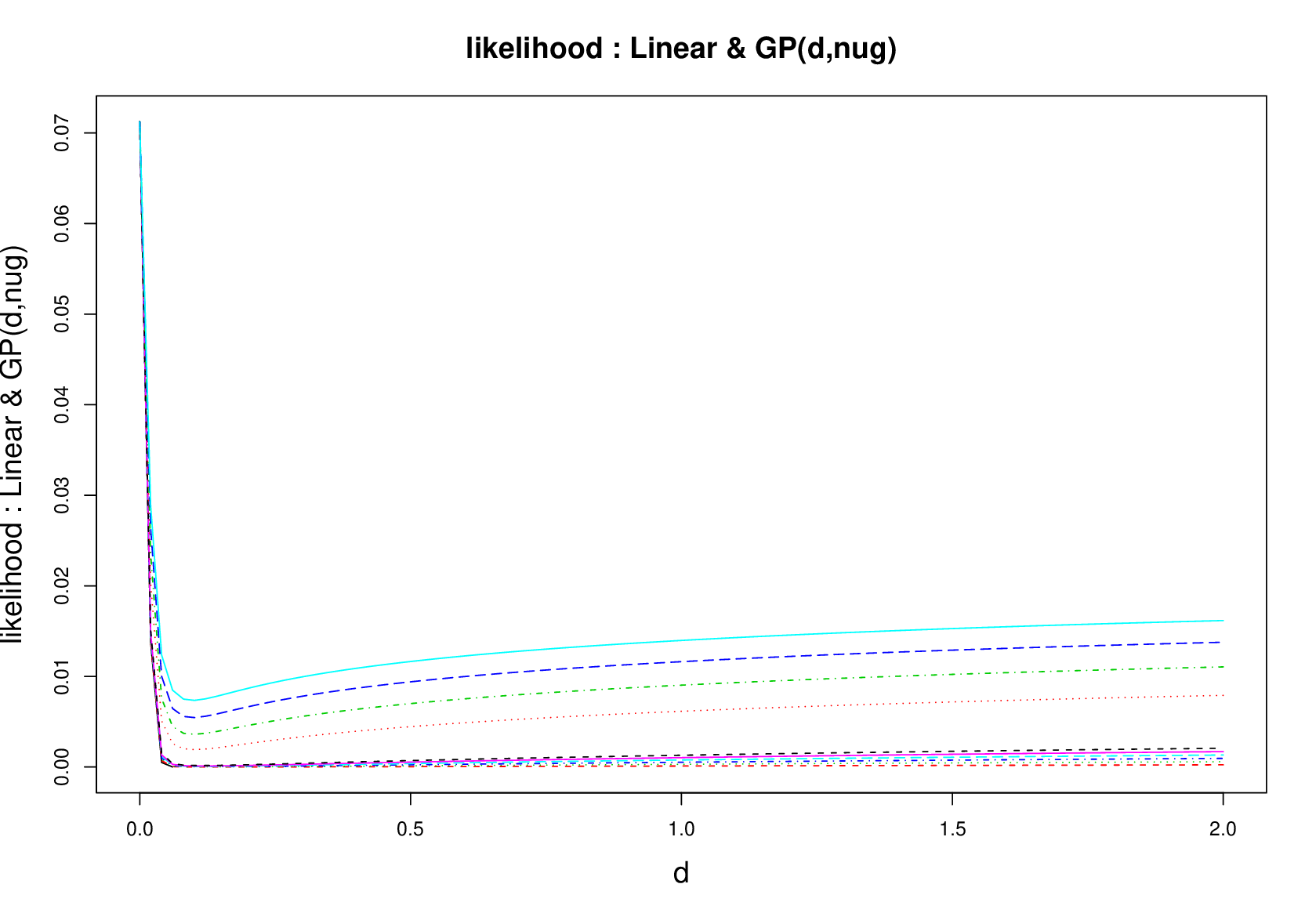} 
\includegraphics[scale=0.205,trim=20 50 10 27,angle=-90]{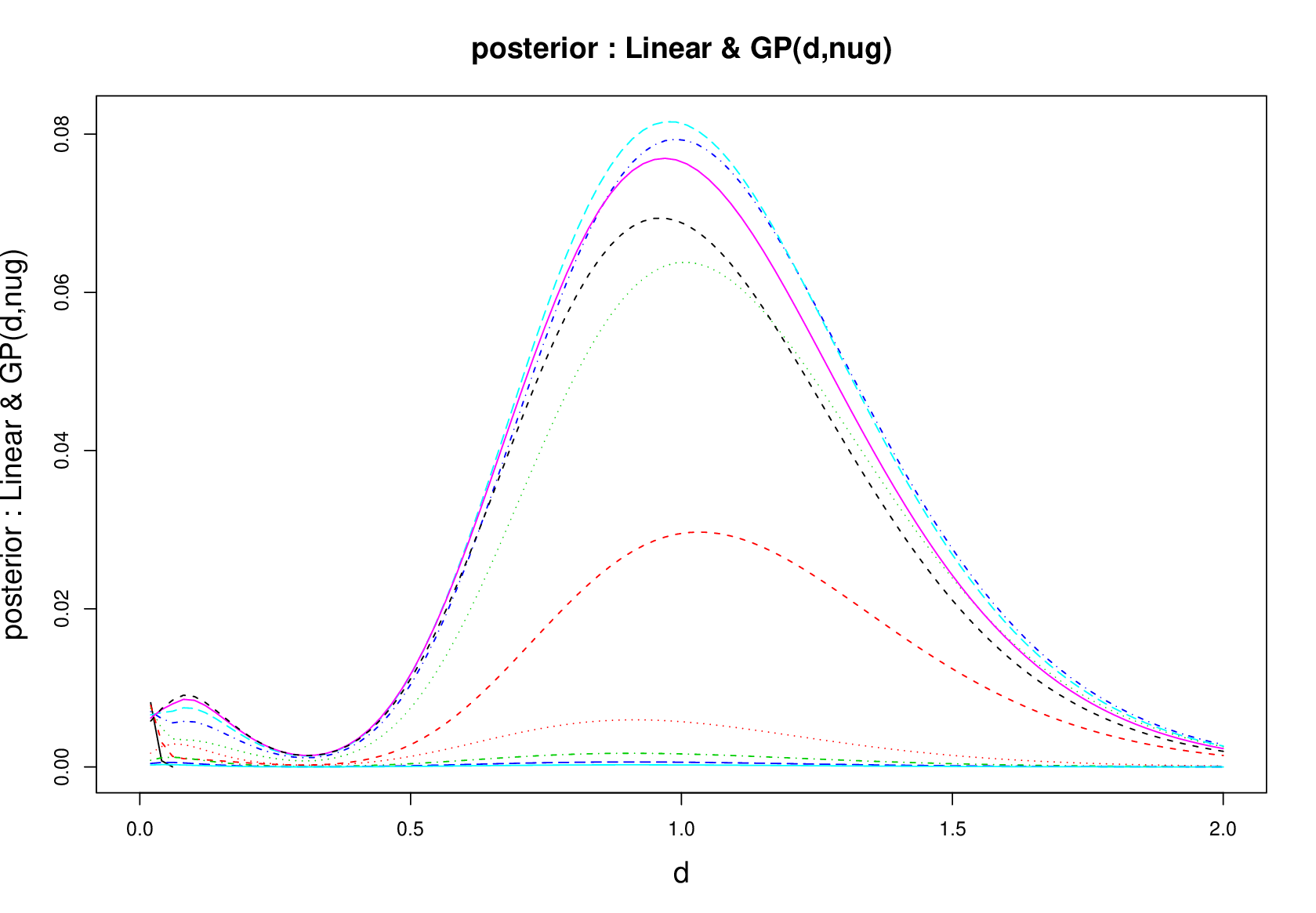} 
\includegraphics[scale=0.205,trim=20 20 10 27,clip=TRUE,angle=-90]{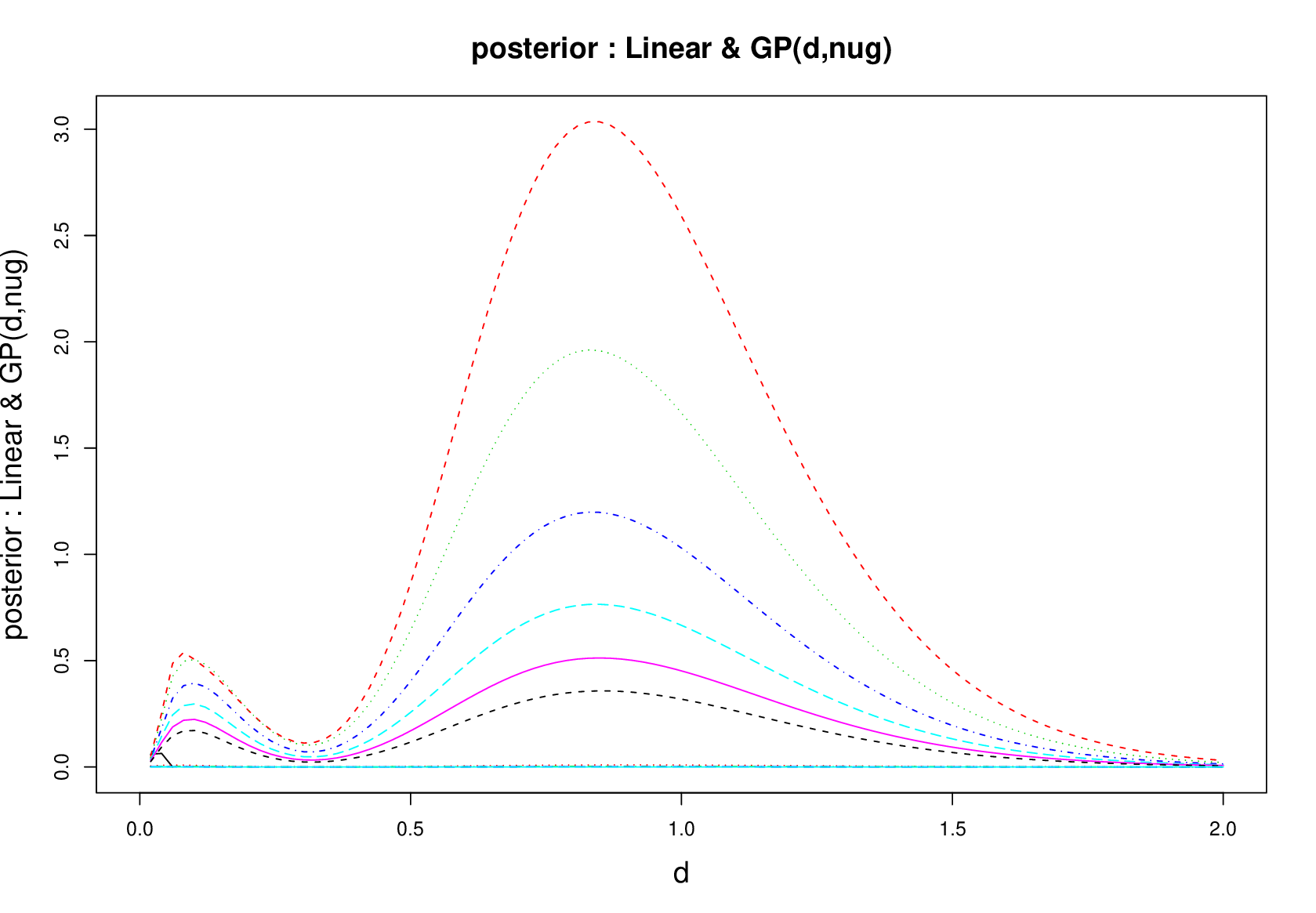}
\includegraphics[scale=0.205,trim=20 20 10 27,clip=TRUE,angle=-90]{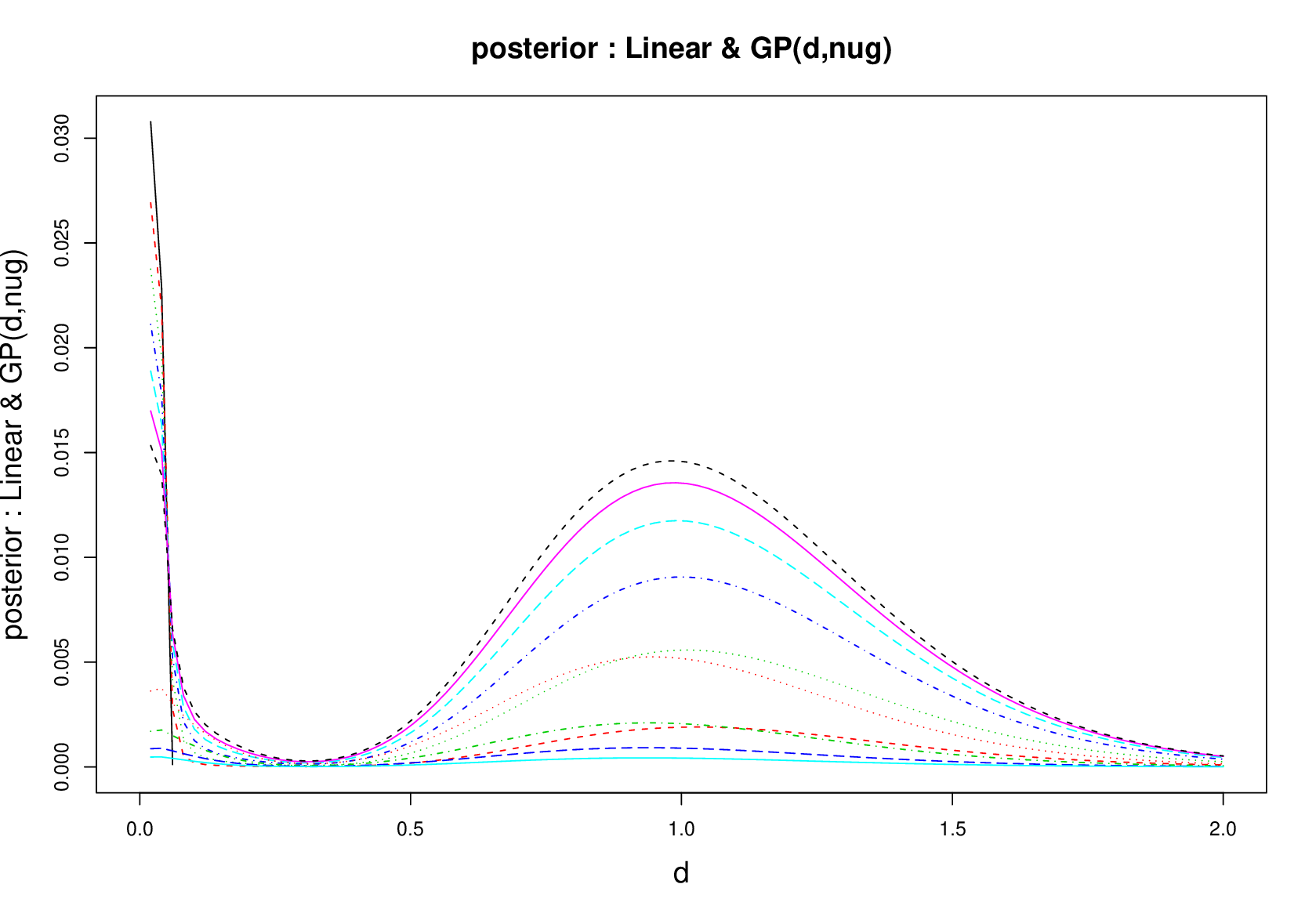}
\caption[Posteriors and likelihoods for linear data] {{\em Top row}
  shows the GP$(d,g)$ fits; {\em Middle row} shows likelihoods and
  {\em bottom row} shows the integrated posterior distribution for
  range ($d$, x-axis) and nugget ($g$, lines) settings for three
  samples, one per each column.  Note that {\tt s2}\;$\equiv
  \hat{\sigma}^2$ in the top row legend(s). }
\label{f:gpvlin:likvpost} 
\end{center} 
\end{figure}

Figure \ref{f:gpvlin:likvpost} shows three samples from the linear
model (\ref{eq:linear:sim}) along with likelihood and posterior
surfaces.  Occasionally the likelihood and posterior lines suddenly
stop due to a numerically unstable parameterization \citep{neal:1997}.
The GP fits shown in the top row of the figure are based on the
maximum {\em a posteriori} (MAP) estimates of $d$, $g$, and
$\sigma^2$.  The posteriors in the bottom row clearly show the
influence of the prior. Nevertheless, the posterior density for large
$d$-values is disproportionately high relative to the prior.  Large
$d$-values represent at least 90\% of the cumulative posterior
distribution.  Samples from these posteriors would yield mostly linear
predictive surfaces.  The last sample is particularly interesting as
well as being the most representative.  Here, the LLM $(d=0)$ is the
MAP GP parameterization, and uniformly dominates all other
parameterizations in posterior density.  Still, the cumulative
posterior density favors large $d$-values, thus favoring linear
``looking'' predictive surfaces over the actual (limiting) linear
parameterization.

\begin{figure}[ht!] 
\begin{center} 
\begin{tabular}{lcr}
\includegraphics[scale=0.24,angle=-90]{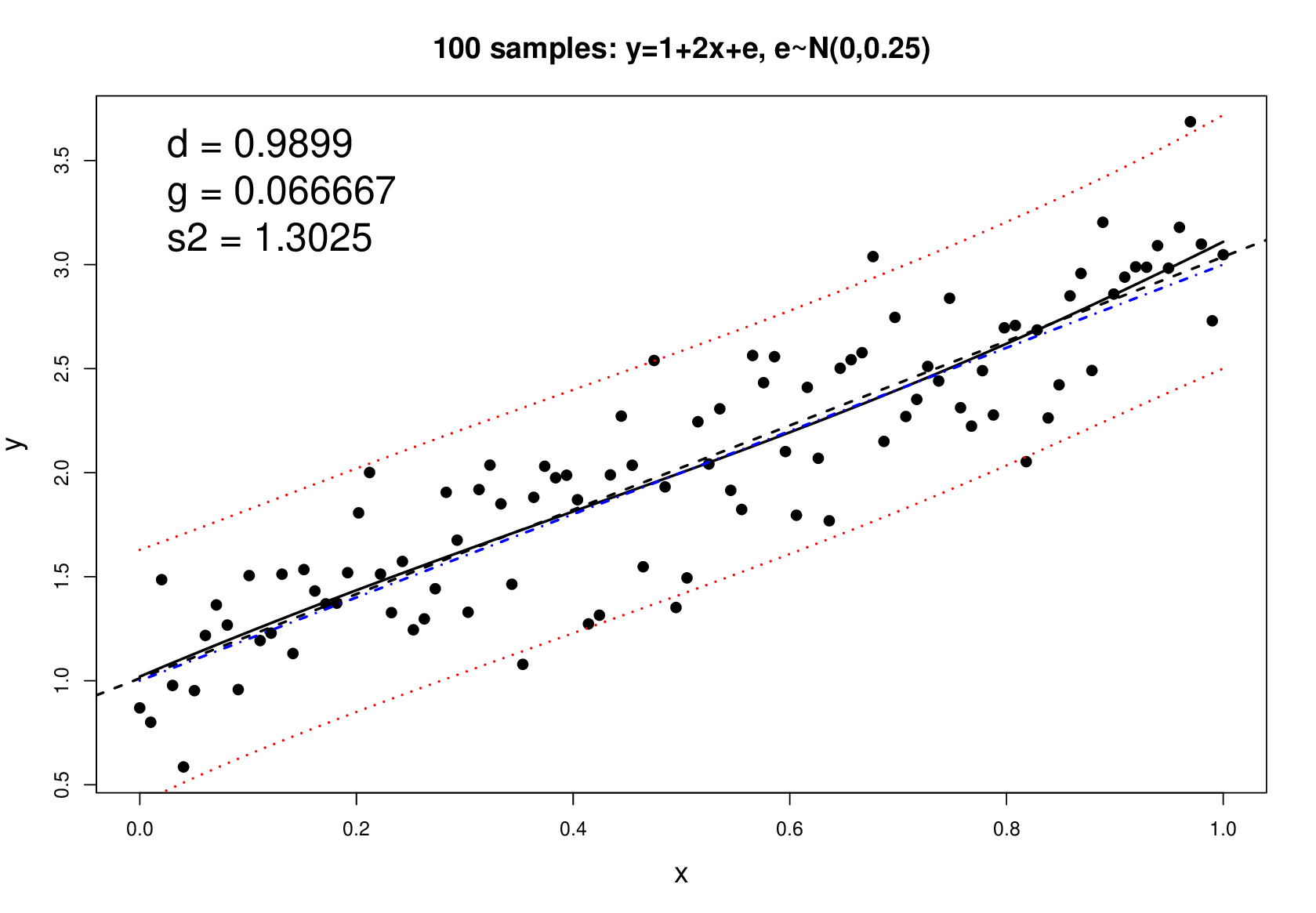} \hfill
\includegraphics[scale=0.24,angle=-90]{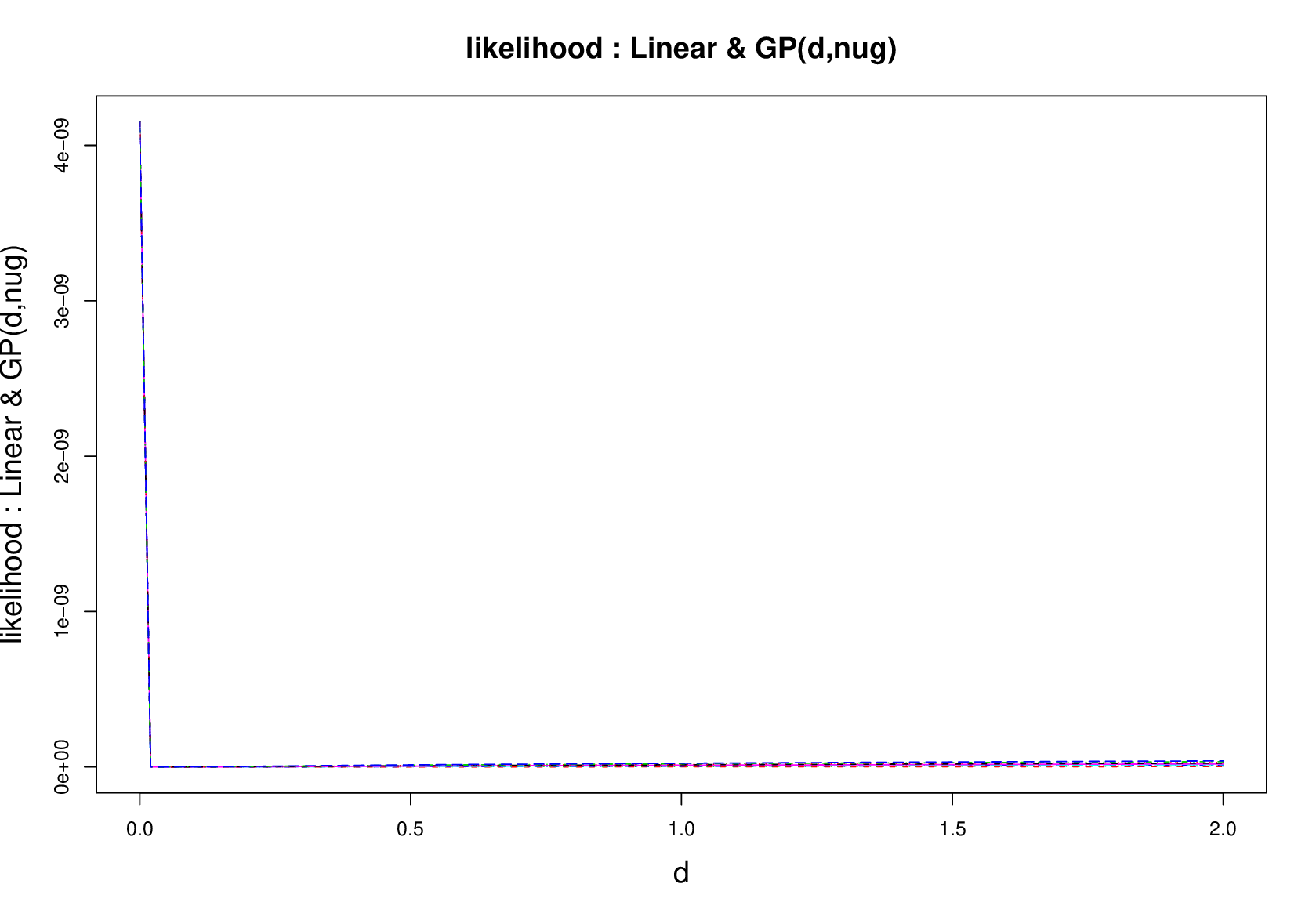} \\
\hfill \includegraphics[scale=0.24,angle=-90]{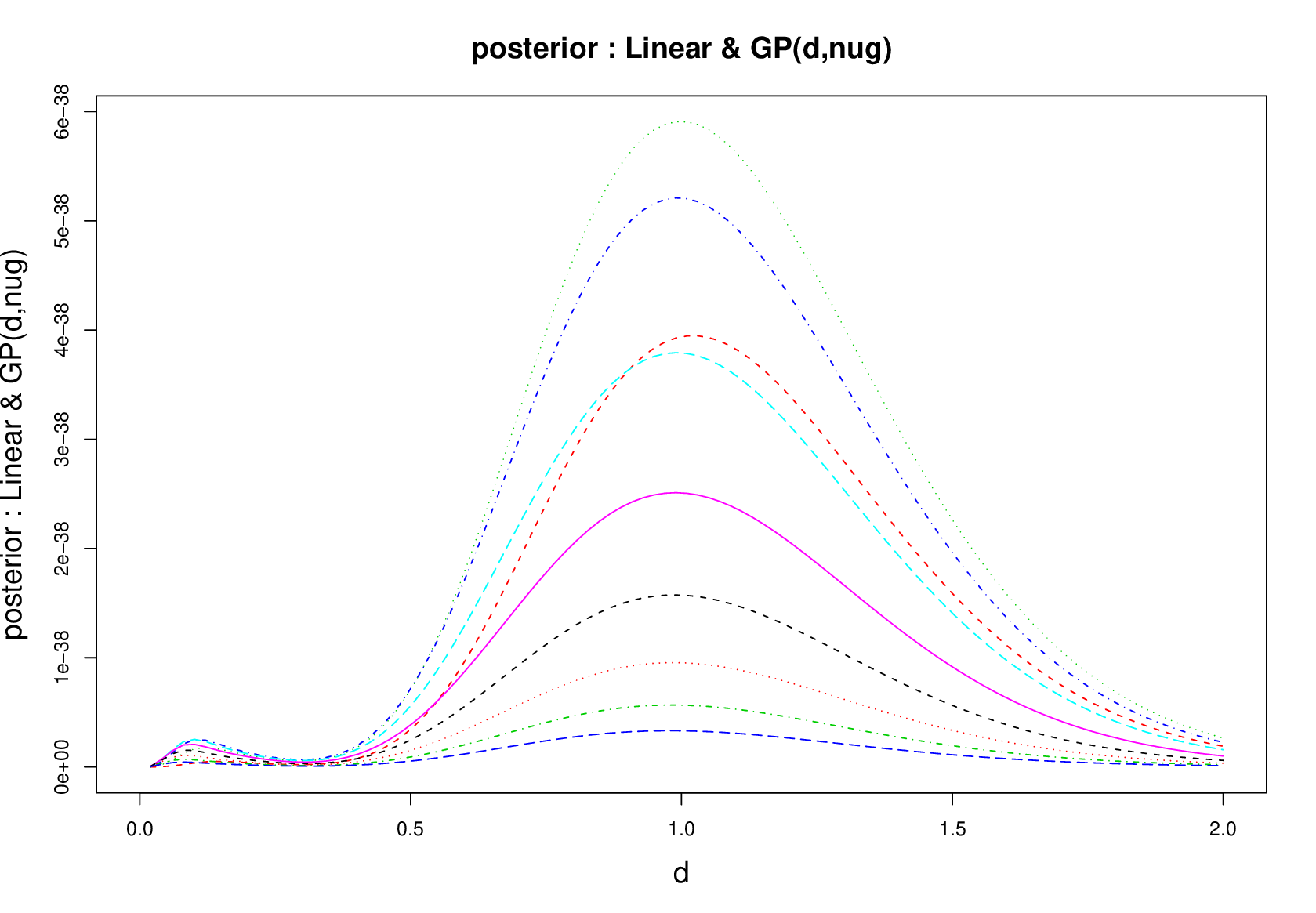} 
\end{tabular}
\caption[GP likelihoods and posteriors for a larger sample]
{{\em Top-left} shows the GP$(d,g)$
fit with a sample of size $n=100$; 
{\em top-right} shows the likelihood and {\em bottom-right}
shows the integrated posterior for range ($d$, x-axis) and
nugget ($g$, lines) settings.}
\label{f:gpvlin:likvpost:n100} 
\end{center} 
\end{figure}

Figure \ref{f:gpvlin:likvpost:n100} ({\em top left}) shows a
representative MAP GP fit for a sample of size $n=100$ from
(\ref{eq:linear:sim}).  Since larger samples have a lower probability
of coming out wavy, the likelihood of the LLM (on the {\em right}) is
much higher than other GP parameterizations, though it is severely
peaked.  Small, nonzero, $d$-values have extremely low likelihood. By
contrast, the posterior in the {\em bottom} panel puts high density on
large $d$-values.  The MAP predictive surface ({\em top left}) has a
very small, but noticeable, amount of curvature.
Ideally, linear looking predictive surfaces should not have to bear
the computational burden implied by full--fledged GPs.  But since the
LLM $(d=0)$ is a point-mass (which is the only parameterization that
actually gives an identity covariance matrix), it has zero probability
under the posterior.  It would never be sampled in an MCMC, even when
it did happen to be the MAP estimate.  %Section \ref{sec:model}
%develops a prior on the range parameter ($d$) so that there is high
%posterior probability of ``jumping'' to the LLM whenever $d$ is
%large.  %The goal is to do
%this without actually proposing $d=0$.

\subsubsection{GP posteriors and likelihoods on non-linear data}
\label{sec:gpllm:gpnonlin}

\begin{figure}[ht!] 
\begin{center} 
%\begin{tabular}{lcr}
%\includegraphics[scale=0.22,trim=0 0 0 0]{plots/linvgp_post_fits_quadcent_2} &
%\includegraphics[scale=0.22,trim=0 0 0 0]{plots/linvgp_lik_dg_quadcent_2} &
%\includegraphics[scale=0.22,trim=0 0 0 0]{plots/linvgp_post_dg_quadcent_2}\\
\includegraphics[scale=0.205,trim=20 50 10 27,angle=-90]{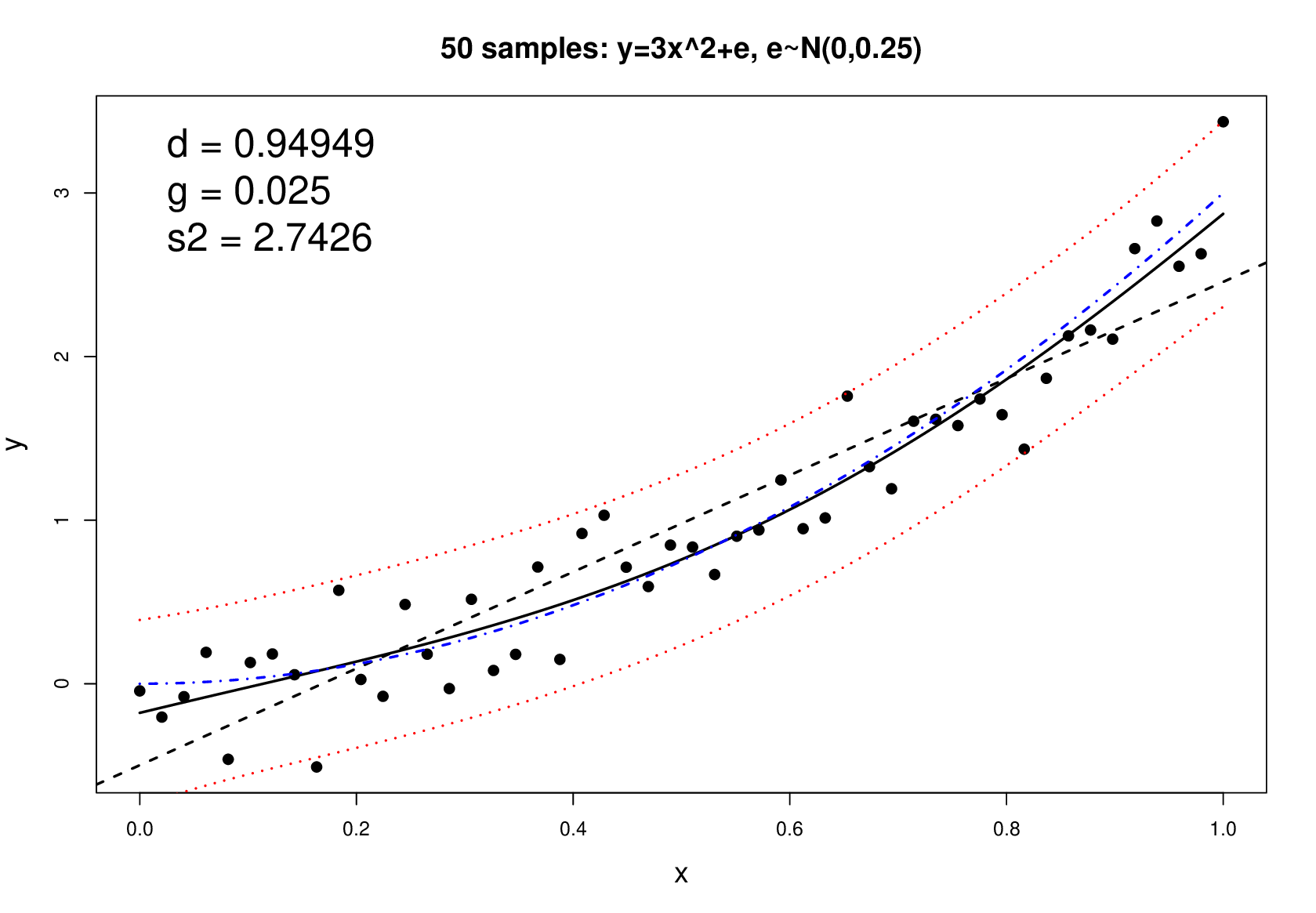} 
\includegraphics[scale=0.205,trim=20 20 10 27,clip=TRUE,angle=-90]{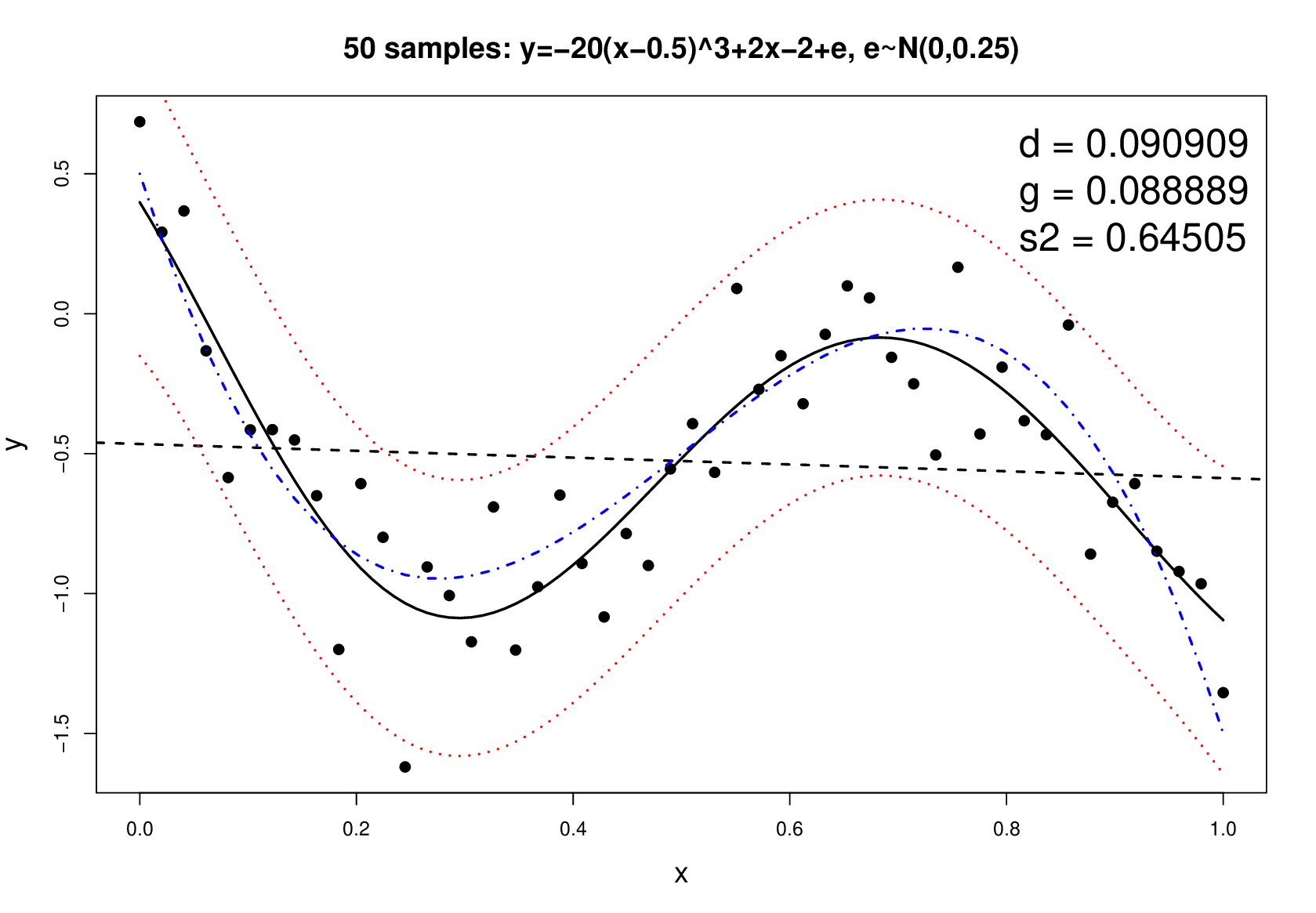} 
\includegraphics[scale=0.205,trim=20 20 10 27,clip=TRUE,angle=-90]{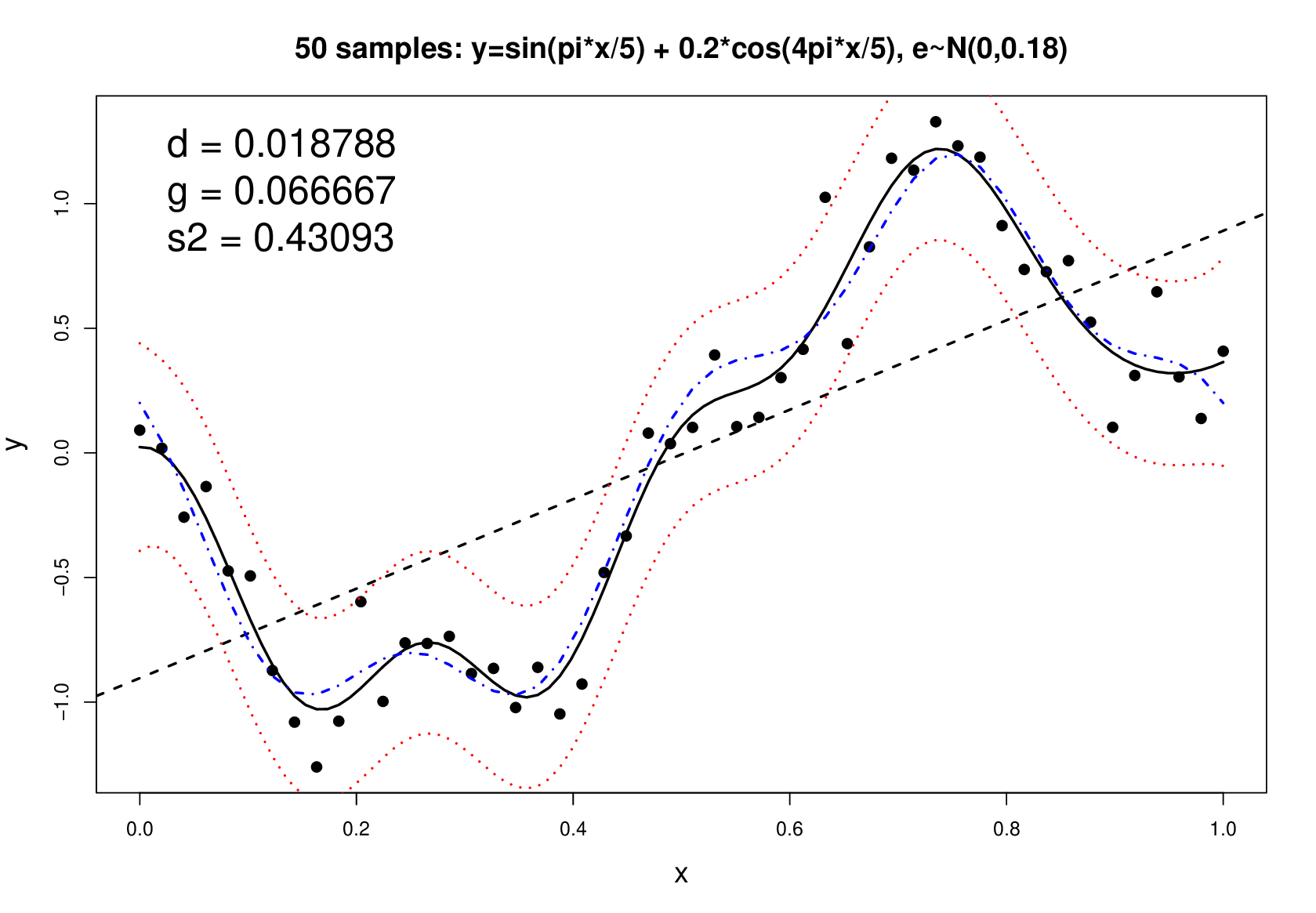} 
\includegraphics[scale=0.205,trim=20 50 10 27,angle=-90]{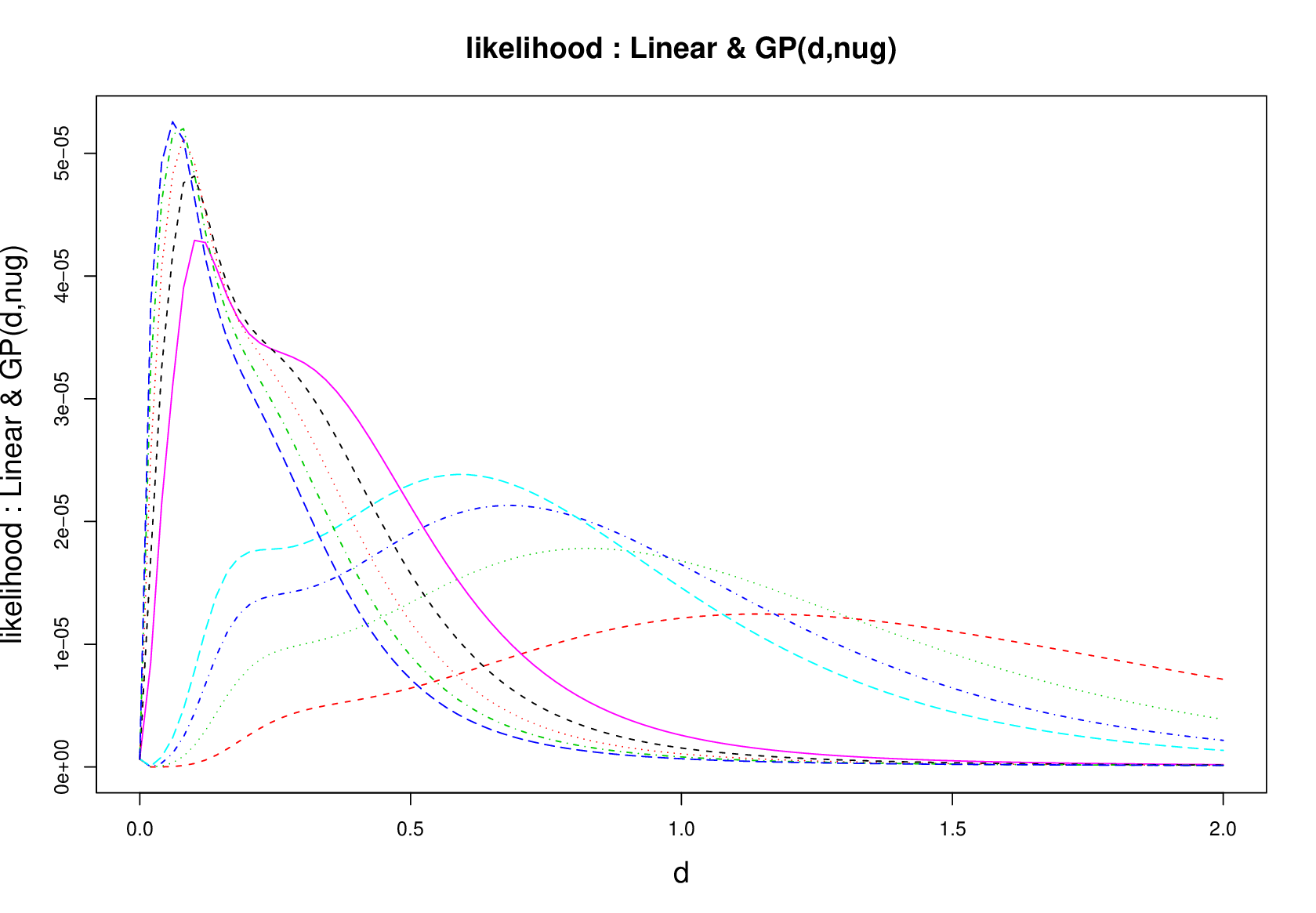}
\includegraphics[scale=0.205,trim=20 20 10 27,clip=TRUE,angle=-90]{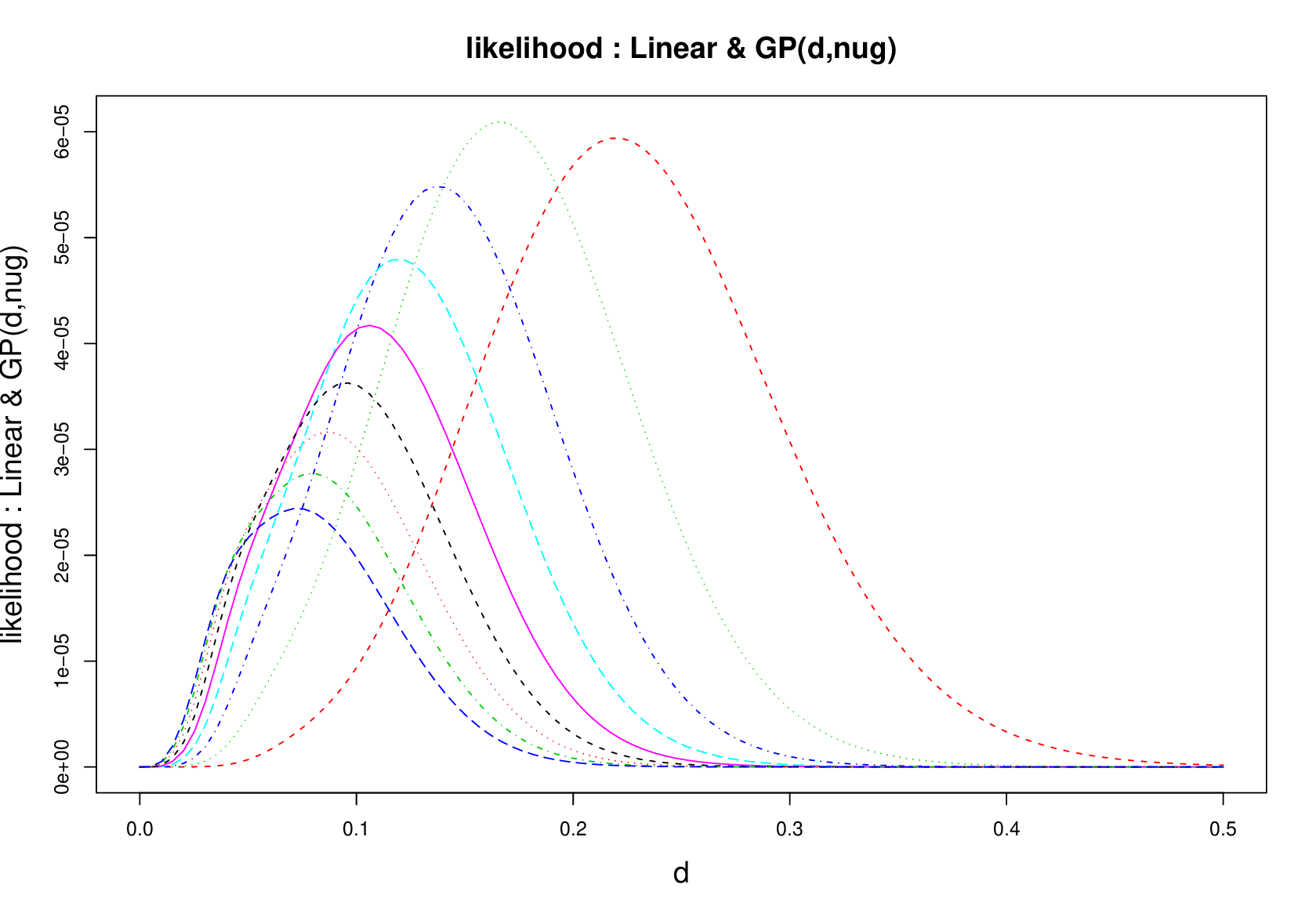} 
\includegraphics[scale=0.205,trim=20 20 10 27,clip=TRUE,angle=-90]{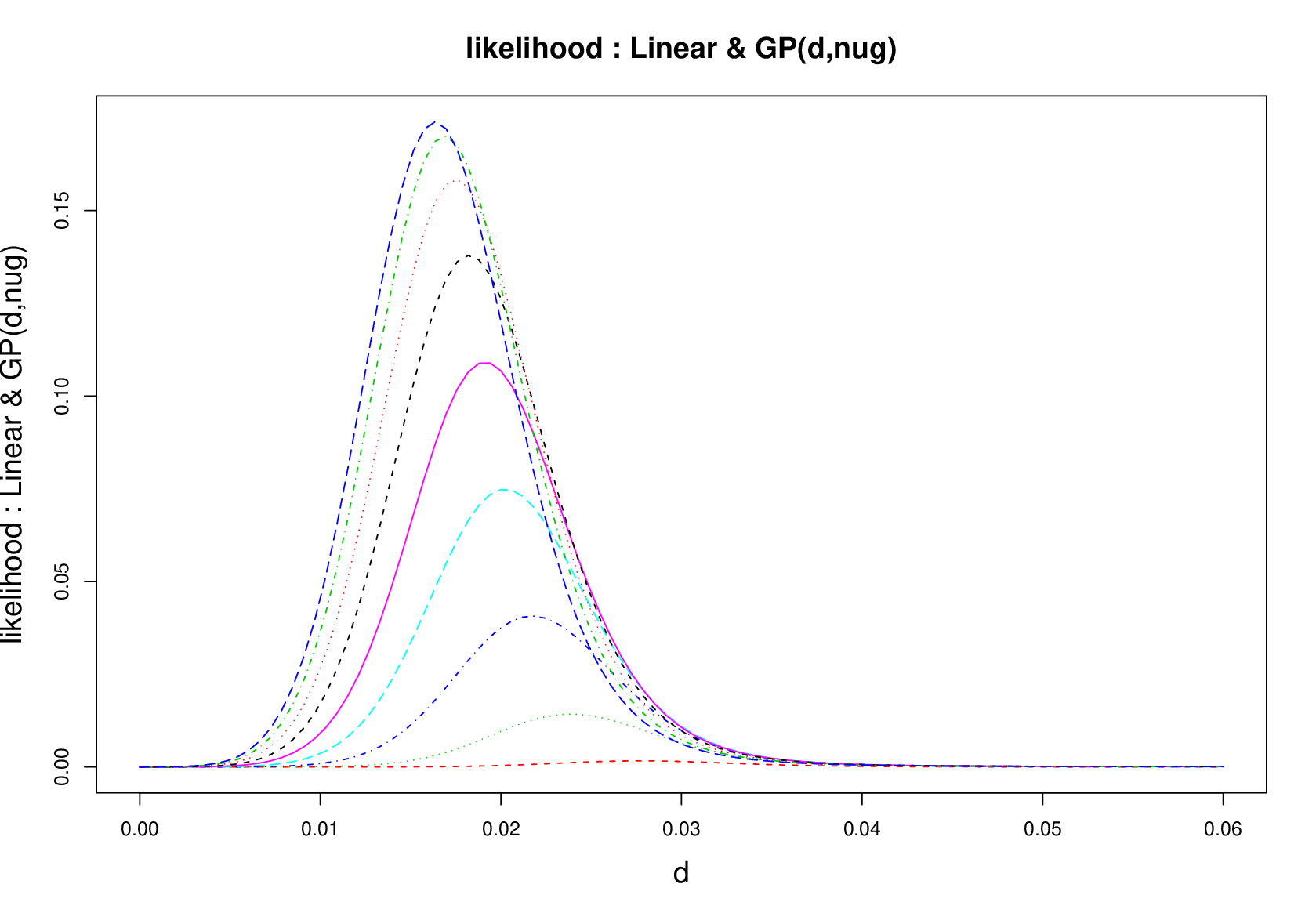}
\includegraphics[scale=0.205,trim=20 50 10 27,angle=-90]{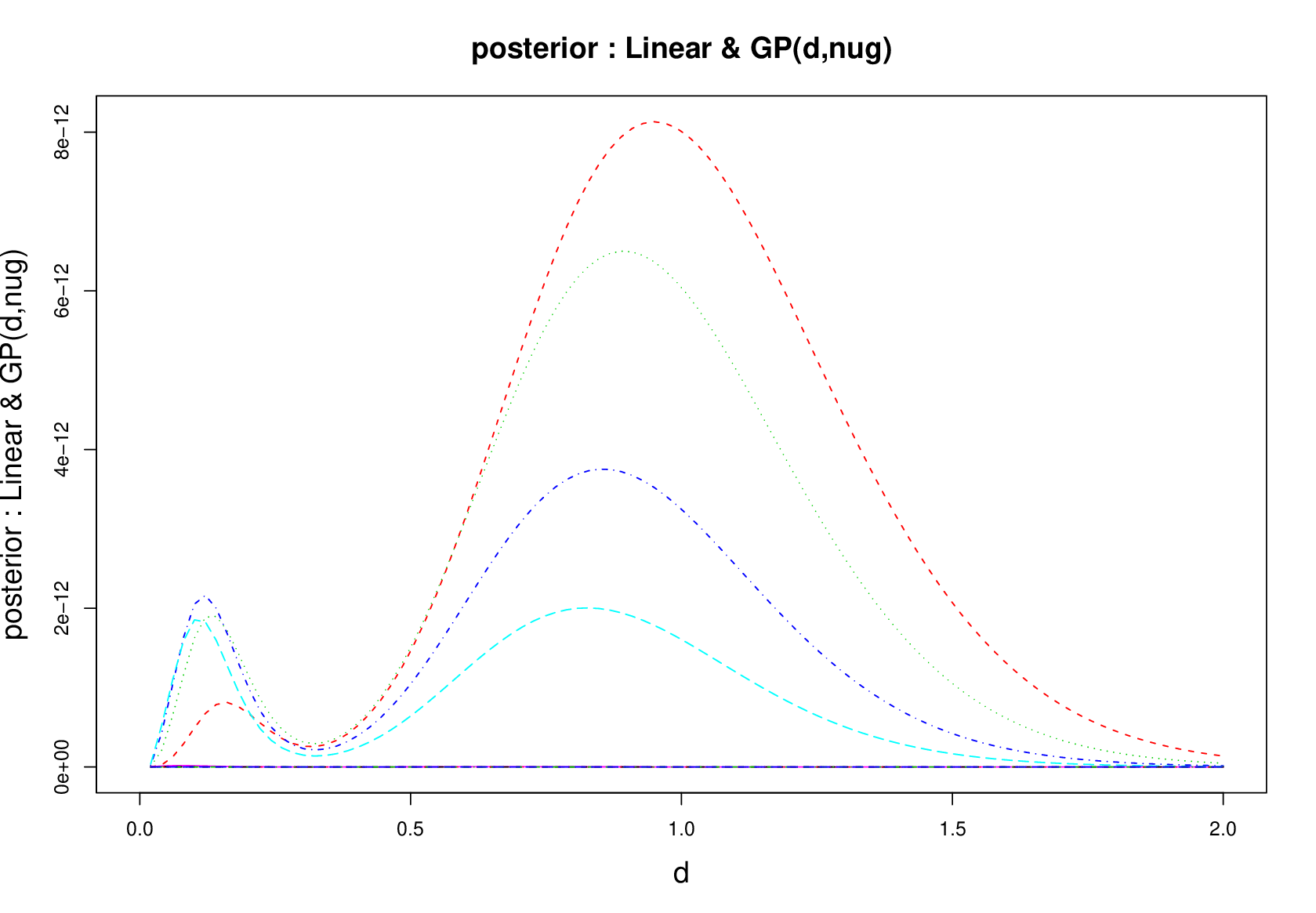}
\includegraphics[scale=0.205,trim=20 20 10 27,clip=TRUE,angle=-90]{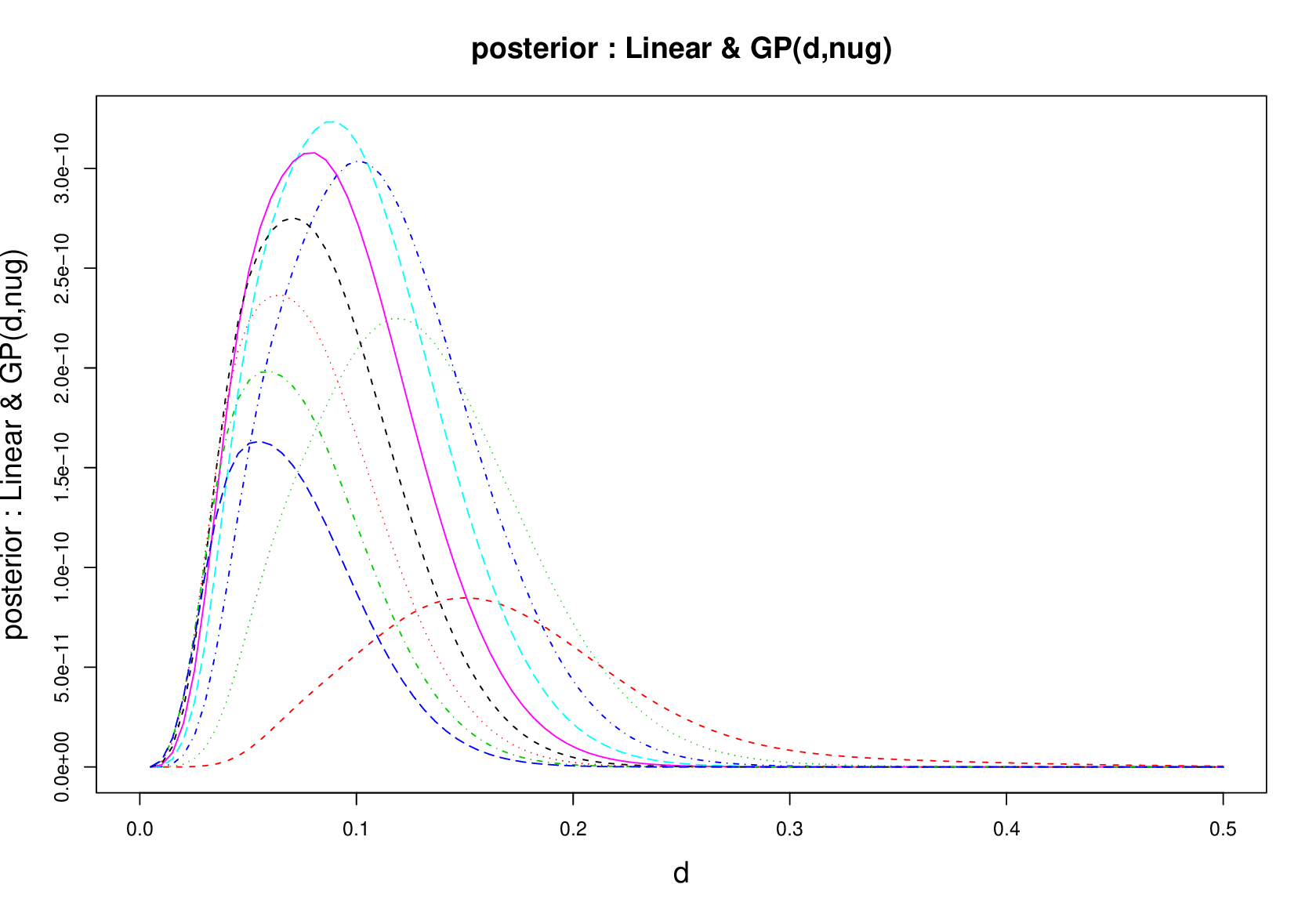}
\includegraphics[scale=0.205,trim=20 20 10 27,clip=TRUE,angle=-90]{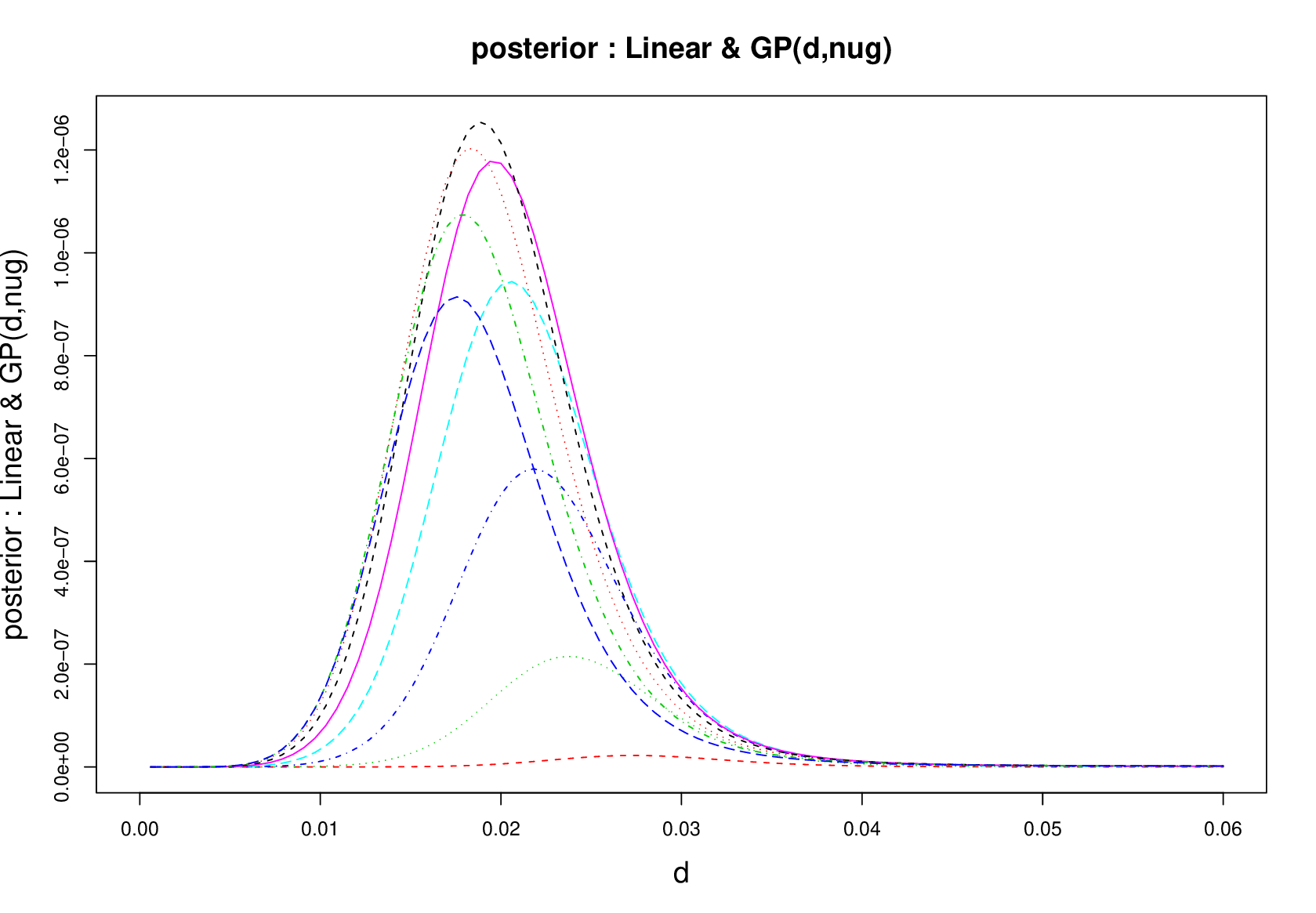}
%\includegraphics[scale=0.22,trim=0 35 0 0]{plots/linvgp_post_fits_linear_2} &
%\includegraphics[scale=0.22,trim=0 35 0 0]{plots/linvgp_lik_dg_linear_2} &
%\includegraphics[scale=0.22,trim=0 35 0 0]{plots/linvgp_post_dg_linear_2}
%\includegraphics[scale=0.22,trim=0 0 0 0]{plots/linvgp_post_fits_exp_2} &
%\includegraphics[scale=0.22,trim=0 0 0 0]{plots/linvgp_lik_dg_exp_2} &
%\includegraphics[scale=0.22,trim=0 0 0 0]{plots/linvgp_post_dg_exp_2} \\
%\end{tabular}
\caption[GP fits on wavy data, Part I]{{\em Top row} shows the
  GP$(d,g)$ fits; {\em Middle row} shows likelihoods and {\em bottom
    row} shows the integrated posterior distribution for range ($d$,
  x-axis) and nugget ($g$, lines) settings for four samples, one per
  each column .As the samples become less linear the $d$-axis (x-axis)
  shrinks in order to focus in on the mode.  }
\label{f:gpvlin:likvpost:wavy1}
\end{center} 
\end{figure}

For completeness, Figure \ref{f:gpvlin:likvpost:wavy1} and shows fits,
likelihoods, and posteriors on non-linear data.  The first column of
Figure \ref{f:gpvlin:likvpost:wavy1} corresponds to a sample with
quadratic mean, and each successive column corresponds to a sample
which is increasingly wavy.
Each sample is of size $n=50$.  The shape of the prior loses its
influence as the data become more non-linear.  
Although in all three cases the MLEs do not correspond to the
MAP estimates, the corresponding ML and MAP predictive surfaces look
remarkably similar (not shown).  This is probably due to the fact that
the posterior integrates out $\bm{\beta}$ and $\sigma^2$, whereas the
likelihoods were computed with point estimates of these parameters.

\section{Model selection prior}
\label{sec:model}

\begin{figure}[ht!]
\centering
\includegraphics[scale=0.28,angle=-90]{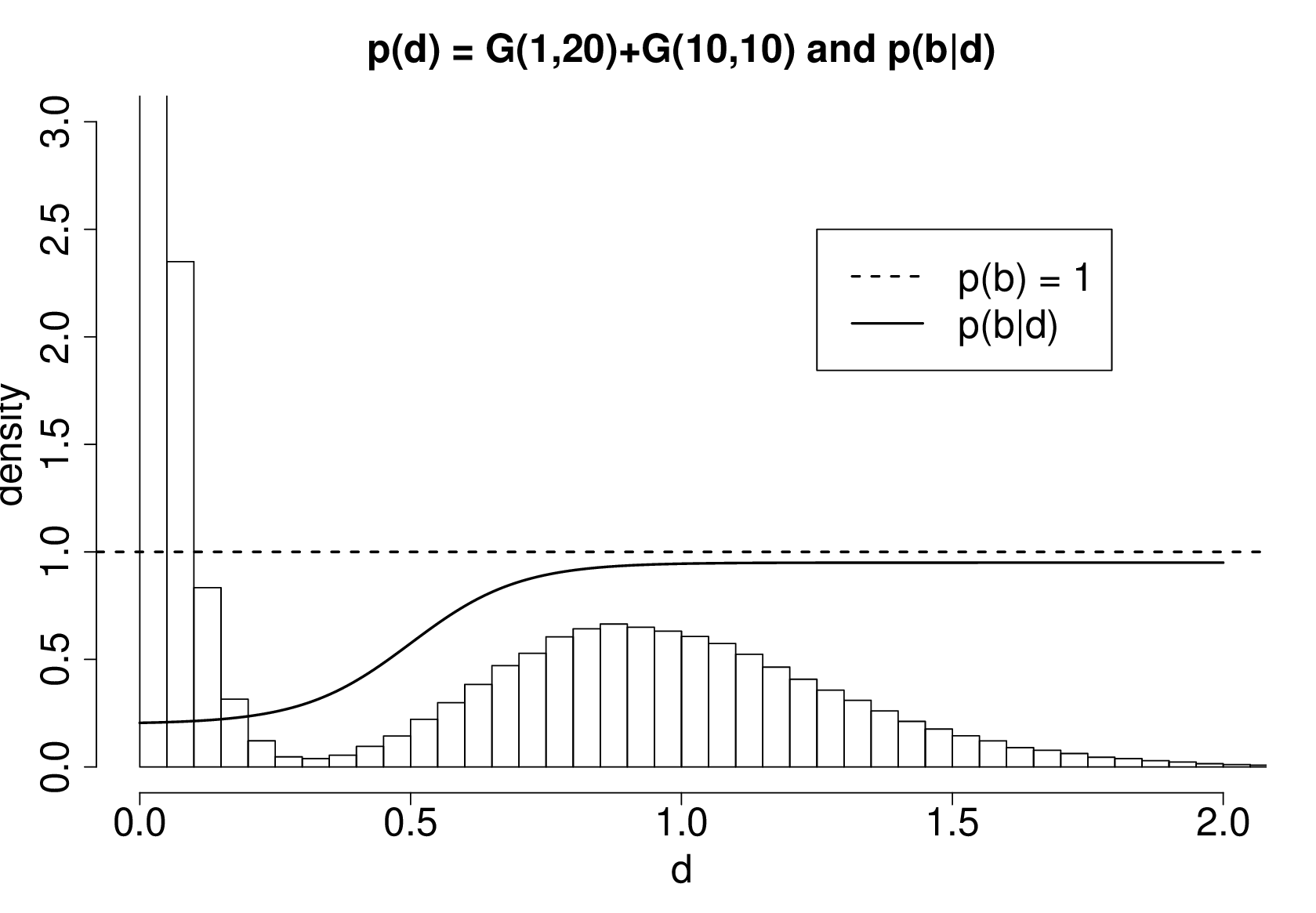}
\caption{Histogram of the mixture of gammas prior $p(d)$ as given in
  Eq.~(\ref{eq:dprior}), with the prior distribution for the boolean ($b$)
  superimposed on $p(d)$ from Eq.(\ref{eq:boolp}) using  
$(\gamma,\theta_1,\theta_2) =(10, 0.2, 0.95)$ .}
\label{f:boolprior}
\end{figure}

With the ideas outlined above, we set out to construct a prior for the
``mixture'' of the GP with its LLM by focusing on large range
parameters rather than $\mb{d} = \mb{0}$ or $g\rightarrow \infty$.
The key idea is an augmentation of the parameter space by $m_X$ latent
indicator variables $\mb{b} = \{b\}_{i=1}^{m_X} \in \{0,1\}^{m_X}$.
The boolean $b_i$ is intended to select either the GP ($b_i=1$) or its
LLM for the $i^{\mbox{\tiny th}}$ dimension.  The actual range
parameter used by the correlation function is multiplied by $\mb{b}$:
e.g. $K^*(\cdot, \cdot| \mb{b} \mb{d})$\footnote{i.e., component--wise
  multiplication---like the ``$\mb{b}$.*$\mb{d}$'' operation in {\tt
    Matlab}}.  To encode our preference that GPs with larger range
parameters should be more likely to ``jump'' to the LLM, the prior on
$b_i$ is specified as a function of the range parameter $d_i$:
$p(b_i,d_i) = p(b_i|d_i)p(d_i)$.

Probability mass functions which increase as a
function of $d_i$, e.g.,  
\begin{equation} 
  p_{\gamma, \theta_1, \theta_2}(b_i=0|d_i) = \theta_1 +
 (\theta_2-\theta_1)/(1 + \exp\{-\gamma(d_i-0.5)\})
\label{eq:boolp}
\end{equation}
with $0<\gamma$ and $0\leq \theta_1 \leq \theta_2 < 1$, can encode
such a preference by calling for the exclusion of dimensions $i$ with
with large $d_i$ when constructing $\mb{K}$.  Thus $b_i$ determines
whether the GP or the LLM is in charge of the marginal process in the
$i^{\mbox{\tiny th}}$ dimension.  Accordingly, $\theta_1$ and
$\theta_2$ represent minimum and maximum probabilities of jumping to
the LLM, while $\gamma$ governs the rate at which $p(b_i=0|d_i)$ grows
to $\theta_2$ as $d_i$ increases.  Figure \ref{f:boolprior} plots
$p(b_i=0|d_i)$ %as in (\ref{eq:boolp})
for $(\gamma,\theta_1,\theta_2) =(10, 0.2, 0.95)$ superimposed on the
mixture of gammas prior $p(d_i)$ from (\ref{eq:dprior}).  The
$\theta_2$ parameter is taken to be strictly less than one so as not
to preclude a GP which models a genuinely nonlinear surface using an
uncommonly large range setting.

The implied prior probability of the full $m_X$-dimensional LLM is
\begin{equation} 
  p(\mbox{linear model}) = \prod_{i=1}^{m_X} p(b_i=0|d_i) 
  = \prod_{i=1}^{m_X} \left[ \theta_1 + \frac{\theta_2-\theta_1}{1 
      + \exp\{-\gamma (d_i-0.5)\}}\right].
\label{e:linp}
\end{equation} 
Observe that the resulting process is still a GP if any of the
booleans $b_i$ are one.  The primary computational advantage
associated with the LLM is foregone unless all of the $b_i$'s are
zero.  However, the intermediate result offers an improvement in
numerical stability in addition to describing a unique transitionary
model lying somewhere between the GP and the LLM.  Specifically, it
allows for the implementation of semiparametric stochastic processes
like $Z(\mb{x}) = \bm{\beta} f(\mb{x}) + \varepsilon(\tilde{\mb{x}})$,
representing a piecemeal spatial extension of a simple linear model.
The first part ($\bm{\beta}f(\mb{x})$) of the process is linear in
some known function of the full set of covariates $\mb{x} =
\{x_i\}_{i=1}^{m_X}$, and $\varepsilon(\cdot)$ is a spatial random
process (e.g., a GP) which acts on a subset of the covariates
$\tilde{\mb{x}}$.  Such models are commonplace in the statistics
community~\citep{dey:1998}.  Traditionally, $\tilde{\mb{x}}$ is
determined and fixed {\em a priori}.  The separable boolean prior in
(\ref{eq:boolp}) implements an adaptively semiparametric process where
the subset $\tilde{\mb{x}} = \{ x_i : b_i = 1, i=1,\dots,m_X \}$ is
given a prior distribution, instead of being fixed.

So even if the computational advantage of the LLM is not present
because some $b_i \ne 0$ we still have that any $b_i = 0$ setting
releases us from the burden of sampling the corresponding $d_i$.  It
also imparts on us the knowledge that the response has (at best) a
linear relationship with the $i^{\mbox{\tiny th}}$ covariate.  This
approach may also increase the scope for analysis of higher
dimensional datasets, where data sparseness in higher dimensions (the
``curse of dimensionality'') can be ameliorated by using linear models
in most dimensions and GP models only in the dimensions where they
will have the most effect, thus reducing the dimension of the GP model
space.  

Note that if an isotropic correlation function is used, which has only
a single range parameter, then only one boolean $b$ is needed, and the
product can be dropped from (\ref{e:linp}).  In this case, however,
the advantage of being able to detect linearity, marginally, in each
dimension is lost.

\subsection{Prediction}
\label{sec:pred}

Prediction under the LLM parameterization of the GP model (\ref{eq:pred})
can be simplified when all of the booleans are zero, whence it is
known that $\mb{K} = (1+g)\mb{I}$.  A characteristic of the standard
linear model is that all input configurations $(\mb{x})$ are treated
as independent conditional on knowing $\bm{\beta}$.  This additionally
implies that in (\ref{eq:pred}) the terms $k(\mb{x})$ and
$K(\mb{x},\mb{x}_j)$ are zero for all $\mb{x}$.  Thus, the predicted
value of $y$ at $\mb{x}$ is normally distributed with mean
$\hat{y}(\mb{x}) = \mb{f}^\top(\mb{x}) \tilde{\bm{\beta}}$ and
variance
\[ \sigma^2 [1 + \tau^2 \mb{f}^\top(\mb{x}) \mb{W} \mb{f}(\mb{x}) -
\tau^2\mb{f}^\top(\mb{x}) \mb{W} \mb{F}^\top ((1+g)\mb{I} +
\tau^2\mb{F}\mb{W}\mb{F}^\top)^{-1} \mb{F}\mb{W}\mb{f}(\mb{x})\tau^2].\]
It is helpful to re-write the above
expression for the variance as
\begin{align}
\hat{\sigma}(\mb{x})^2 &= \label{eq:varhelp}
        \sigma^2 [1 + \tau^2 \mb{f}^\top(\mb{x}) \mb{W} \mb{f}(\mb{x})]\\
&- \sigma^2 \left[1 +
\tau^2 \mb{f}^\top(\mb{x}) \mb{W} \mb{f}(\mb{x}) -
\frac{\tau^2}{1+g}\mb{f}^\top(\mb{x})\mb{W}\mb{F}^\top \left(\mb{I} +
\frac{\tau^2}{1+g}\mb{F}\mb{W}\mb{F}^\top\right)^{-1}
\mb{F}\mb{W}\mb{f}(\mb{x})\tau^2\right]. \nonumber
\end{align}
A matrix inversion lemma called the Woodbury formula
\citep[][pp.~51]{golub:1996} or the Sherman--Morrison--Woodbury
formula \citep[][pp.~67]{berns:2005} states that for $(\mb{I} +
\mb{V}^\top \mb{A} \mb{V})$ non-singular $ (\mb{A}^{-1} +
\mb{V}\mb{V}^\top)^{-1} = \mb{A} - (\mb{A}\mb{V})(\mb{I} +
\mb{V}^\top\mb{A}\mb{V})^{-1} \mb{V}^\top\mb{A}.  $ Taking $\mb{V}
\equiv \mb{F}^\top(1+g)^{-\frac{1}{2}}$ and $\mb{A} \equiv
\tau^2\mb{W}$ in (\ref{eq:varhelp}) gives
\begin{equation} 
\hat{\sigma}(\mb{x})^2 = \sigma^2 \left[1
+ \mb{f}^\top(\mb{x}) \left(\frac{\mb{W}^{-1}}{\tau^2} + \frac{\mb{F}^\top
\mb{F}}{1+g}\right)^{-1} \mb{f}(\mb{x})\right]. 
\label{eq:lin:predvar:simple}
\end{equation} 
Eq.~(\ref{eq:lin:predvar:simple}) is not only a simplification of the
predictive variance given in (\ref{eq:pred}), but it should look
familiar.  Writing $\mb{V}_{\tilde{\beta}}$ with $\mb{K}^{-1} =
\mb{I}/(1+g)$ in (\ref{eq:betavar}) gives
\begin{equation}
\mb{V}_{\tilde{\beta}} =
\left(\frac{\mb{W}^{-1}}{\tau^2} + \frac{\mb{F}^\top
    \mb{F}}{1+g}\right)^{-1}
\;\;\;\;\; \mbox{and then:} \;\;\;\;\; 
\hat{\sigma}(\mb{x})^2 = \sigma^2 \left[1 +
\mb{f}^\top(\mb{x})\mb{V}_{\tilde{\beta}} \mb{f}(\mb{x})\right].
\label{eq:vb:linear} 
\end{equation} 
This is just the usual posterior predictive density at $\mb{x}$ under
the standard linear model: $y(\mb{x})~\sim~N[\mb{f}^\top(\mb{x})
\tilde{\bm{\beta}}, \sigma^2(1 + \mb{f}^\top(\mb{x})
\mb{V}_{\tilde{\beta}} \mb{f}(\mb{x}))]$.  This means that we have a
choice when it comes to obtaining samples from the posterior
predictive distribution under the LLM.  We prefer (\ref{eq:vb:linear})
over (\ref{eq:pred}) because the latter involves inverting the
$n\times n$ matrix $\mb{I} + \tau^2\mb{F}\mb{W}\mb{F}^\top/(1+g)$,
whereas the former only requires the inversion of an $m \times m$
matrix.  Of course, if any of the booleans are non-zero, then GP
prediction must proceed as usual, following Eq.~(\ref{eq:pred}).

GP fits to typical nonlinear response surfaces may seldom achieve $b_i
= 0$ for all $i$.  However, the following section illustrates how
treed partitioning can dramatically increase the probability of
jumping to the LLM parameterization (in at least part of the input
space) where the more thrifty predictive equations
(\ref{eq:vb:linear}) may be exploited.

\section{Implementation, results, and comparisons}
\label{sec:results}

Here, the GP with jumps to the LLM (hereafter GP LLM) is illustrated
on synthetic and real data.  This work grew out of research focused on
extending the reach of the treed GP model presented by
\citet{gra:lee:2008}, whereby the data are recursively partitioned and
a separate GP is fit in each partition.  Thus most of our experiments
are in this context, though in Section~\ref{sec:fried} we demonstrate
an example without treed partitioning.  Partition models are an ideal
setting for evaluating the utility of the GP LLM, as linearity can be
extracted in large areas of the input space.  The result is a uniquely
tractable nonstationary semiparametric spatial model.

A separable correlation function is used throughout this section for
brevity and consistency, even though in some cases the process which
generated the data is clearly isotropic.  Proposals for the booleans
$\mb{b}$ are drawn from the prior, conditional on $\mb{d}$, and
accepted and rejected on the basis of the constructed covariance
matrix $\mb{K}$.  The same prior parameterizations are used for all
experiments unless otherwise noted, the idea being to develop a method
that works ``right out of the box'' as much as possible.

\subsection{Synthetic exponential data}
\label{sec:exp}

Consider the 2-d input space $[-2,6] \times [-2,6]$ in which the true
response is given by $ Y(\mathbf{x}) = x_1 \exp(-x_1^2 - x_2^2) +
\epsilon$, where $\epsilon \sim N(0,\sigma=0.001)$.  
\begin{figure}[ht!] 
\vspace{0.4cm}
\begin{center}
\includegraphics[scale=0.3,angle=-90,trim=10 175 75 175]{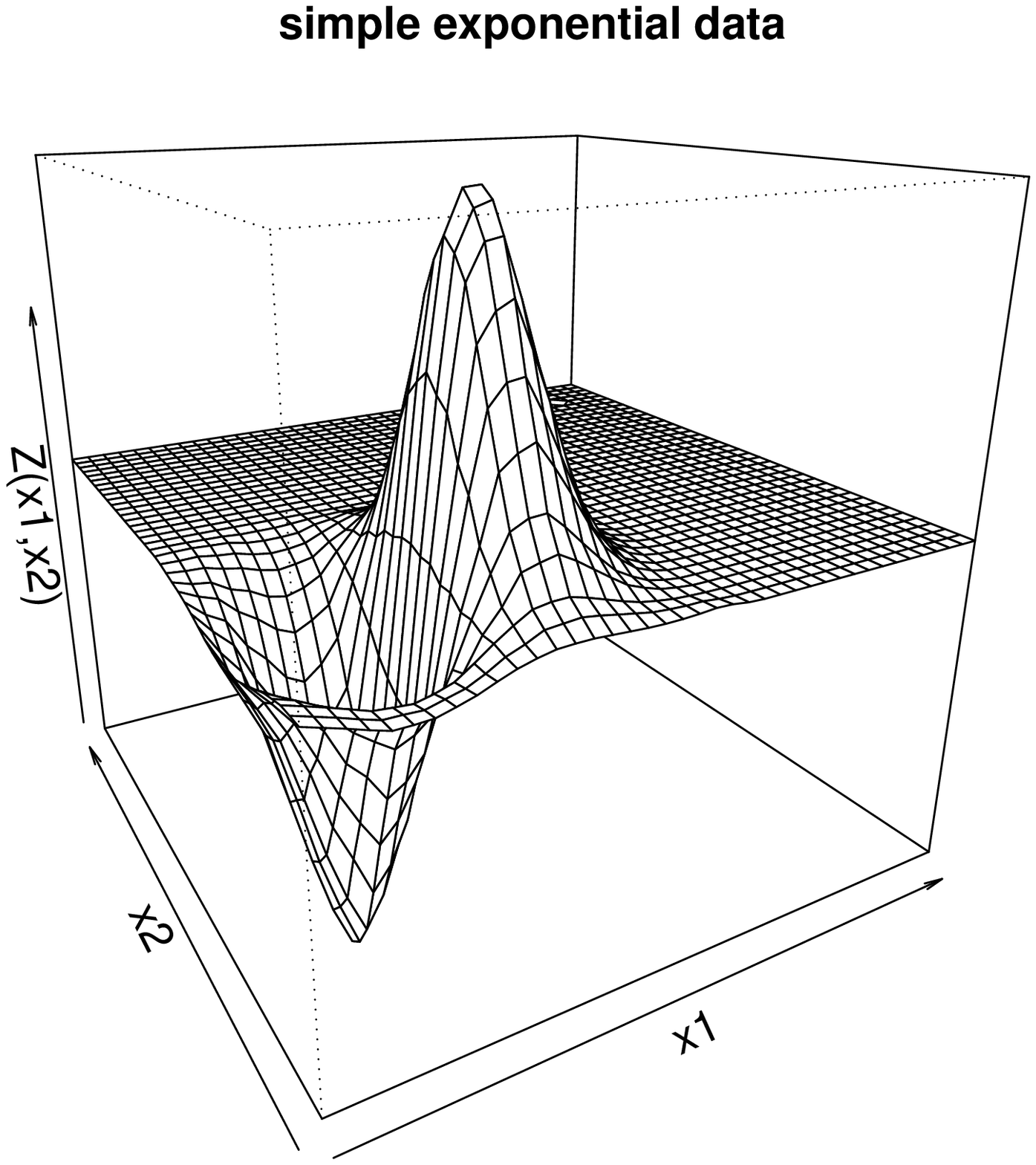} 
\hspace{1cm}
\includegraphics[scale=0.27,angle=-90]{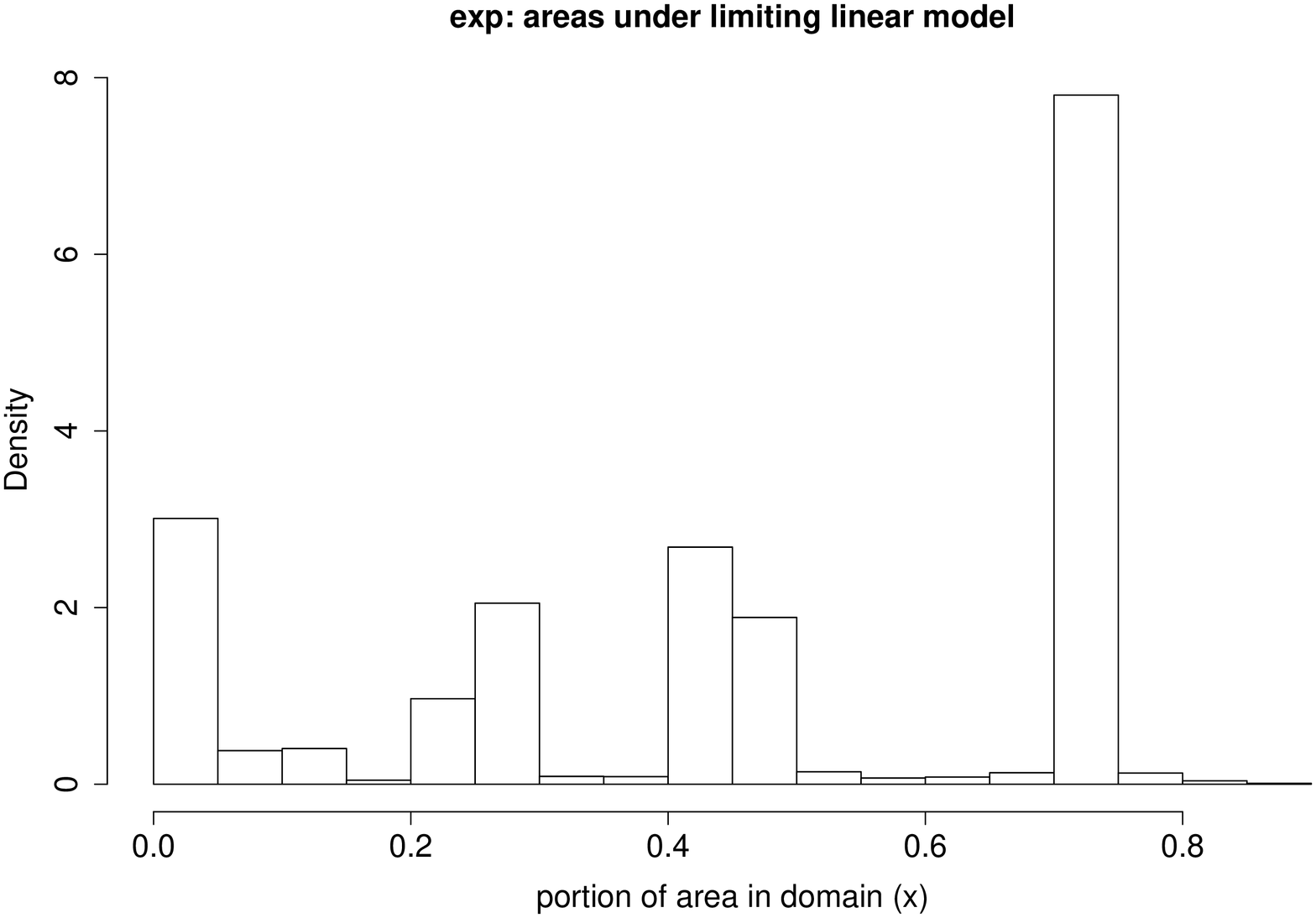}
\caption{ {\em Left:} exponential data GP LLM fit.
{\em Right:} histogram of the areas under the LLM.}
\label{f:lin:exp} 
\end{center}
\end{figure}
Figure \ref{f:lin:exp} summarizes the consequences of estimation and
prediction with the treed GP LLM for a $n=200$ random sub-sample of
this data from a regular grid of size 441.  The partitioning structure
of the treed GP LLM first splits the region into two halves, one of
which can be fit linearly.  It then recursively partitions the half
with the action into a piece which requires a GP and another piece
which is also linear.  The {\em left} pane shows a mean predictive
surface wherein the LLM was used over 66\% of the domain (on average)
which was obtained in less than ten seconds on a 1.8 GHz Athalon.  The
{\em right} pane shows a histogram of the areas of the domain under
the LLM over 20-fold repeated experiments.  The four modes of the
histogram clump around 0\%, 25\%, 50\%, and 75\% showing that most
often the obvious three--quarters of the space is under the LLM,
although sometimes one of the two partitions will use a very smooth
GP.  The treed GP LLM was 40\% faster than the treed GP alone when
combining estimation and sampling from the posterior predictive
distributions at the remaining $n'=241$ points from the grid.

\subsection{Motorcycle Data}
\label{sec:moto}

\begin{figure}[ht!]
\centering
\includegraphics[angle=-90,scale=0.28]{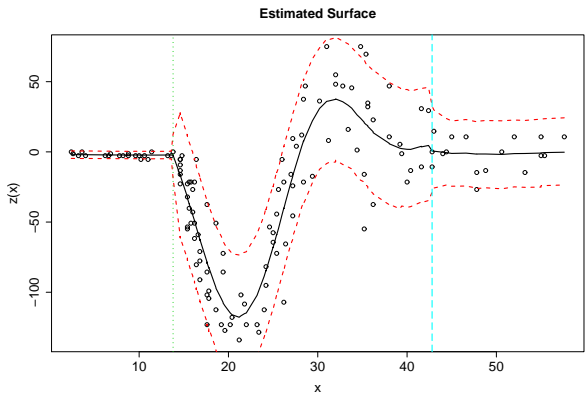}
\caption{Motorcycle Data fit by treed GP LLM.}
\label{f:lin:moto}
\end{figure}

The Motorcycle Accident Dataset \citep{silv:1985} is a classic for
illustrating nonstationary models.  It samples the acceleration force
on the head of a motorcycle rider as a function of time in the first
moments after an impact.  Figure \ref{f:lin:moto} shows the data and a
fit using the treed GP LLM.  The plot shows the mean predictive
surface with 90\% quantile error bars, along with a typical partition.
On average, 29\% of the domain was under the LLM, split between the
left low--noise region (before impact) and the noisier rightmost
region.

\citet{rasm:ghah:nips:2002} analyzed this data using a Dirichlet
process mixture of Gaussian process (DPGP) experts which reportedly
took one hour on a 1 GHz Pentium.  Such times are typical of inference
under nonstationary models because of the computational effort
required to construct and invert large covariance matrices.  In
contrast, the treed GP LLM fits this dataset with comparable accuracy
but in less than one minute on a 1.8 GHz Athalon.

We identify three things which make the treed GP LLM so fast relative
to most nonstationary spatial models.  (1) Partitioning fits models to
less data, yielding smaller matrices to invert.  (2) Jumps to the LLM
mean fewer inversions all together.  (3) MCMC mixes better because
under the LLM the parameters $\mb{d}$ and $g$ are out of the picture
and all sampling can be performed via Gibbs steps.

\subsection{Friedman data}
\label{sec:fried}

This Friedman dataset is the first one of a suite that was used to
illustrate MARS (Multivariate Adaptive Regression Splines)
\citep{freid:1991}.  There are 10 covariates in the data ($\mb{x} =
\{x_1,x_2,\dots,x_{10}\}$), but the function that describes the
responses ($Y$), observed with standard Normal noise,
\begin{equation}
E(Y|\mb{x}) = \mu = 10 \sin(\pi x_1 x_2) + 20(x_3 - 0.5)^2 + 10x_4 + 5 x_5
\label{eq:f1}
\end{equation}
depends only on $\{x_1,\dots,x_5\}$, thus combining nonlinear, linear,
and irrelevant effects.  We make comparisons on this data to results
provided for several other models in recent literature.
\citet{chip:geor:mccu:2002} used this data to compare their treed LM
algorithm to four other methods of varying parameterization: linear
regression, greedy tree, MARS, and neural networks.  The statistic
they use for comparison is root mean--squared error, $\mbox{RMSE} =
\sqrt{ \sum_{i=1}^n (\mu_i - \hat{Y}_i)^2/n}$, where $\hat{Y}_i$ is
the model--predicted response for input $\mb{x}_i$.  The $\mb{x}$'s
are randomly distributed on the unit interval.  RMSE's are gathered
for fifty repeated simulations of size $n=100$ from (\ref{eq:f1}).
Chipman et al. provide a nice collection of boxplots showing the
results.  However, they do not provide any numerical results, so we
have extracted some key numbers from their plots and refer the reader
to their paper for the full results.

We duplicated the experiment using both a stationary GP and our GP
LLM.  For this dataset, we use a single model, not a treed model, as
the function is essentially stationary in the spatial statistical
sense (so if we were to try to fit a treed GP, it would keep all of
the data in a single partition).  Linearizing boolean prior parameters
$(\gamma,\theta_1,\theta_2)=(10,0.2,0.9)$ were used, which gave the
LLM a relatively low prior probability of 0.35, for large range
parameters $d_i$.  The RMSEs that we obtained for the stationary GP
and the GP LLM are summarized in the table below.
\begin{center}
\singlespacing
\begin{tabular}{l|rrrrrr}
        & Min & 1st Qu. & Median & Mean & 3rd Qu. & Max \\
\hline
GP LLM  & 0.4341& 0.5743& 0.6233& 0.6258& 0.6707& 0.7891 \\
GP      & 0.8196& 0.8835& 0.9131& 0.9232& 0.9710& 1.0440 \\
Linear  &1.710  &2.165  & 2.291 & 2.325 &2.500  & 2.794 \\
\end{tabular}
\vspace{0.4cm}
\end{center}
Results on the linear model are reported for calibration purposes, and
can be seen to be essentially the same as those reported by Chipman et
al.  RMSEs for the GP LLM are on average significantly better than
{\em all} of those reported for the above methods, with lower
variance.  For example, the best mean RMSE shown in the boxplot is
$\approx0.9$.  That is 1.4 times higher than the worst one we obtained
for GP LLM.  Further comparison to the boxplots provided by Chipman et
al. shows that the GP LLM is the clear winner.  It is also clear that
jumping to a linear model in the relevant dimensions provides a more
stable fit that gives improved predictive performance relative to a
full (stationary) GP.

In fitting the model, the Markov Chain quickly keyed in on the fact
that only the first three covariates contribute nonlinearly.  After
burn--in, the booleans $\mb{b}$ almost never deviated from
$(1,1,1,0,0,0,0,0,0,0)$.  From the following table summarizing the
posterior for the linear regression coefficients ($\bm{\beta}$) we can
see that the coefficients for $x_4$ and $x_5$ (between double-bars)
were estimated accurately, and that the model correctly determined
that $\{x_6,\dots x_{10}\}$ were irrelevant (i.e. not included in the
GP, and had $\beta$'s close to zero).
\begin{center}
\singlespacing
\begin{tabular}{c|c||rr||rrrrr}
   & &  $x_4$ & $x_5$ & $x_6$ & $x_7$ & $x_8$ & $x_9$ & $x_{10}$ \\
 \hline
 & 5\% Qu. & 8.40 & 2.60 & -1.23 & -0.89 & -1.82 & -0.60 & - 0.91\\
 $\bm{\beta}$ & Mean & 9.75 & 4.59 & -0.190 & 0.049 & -0.612 & 0.326 & 0.066 \\
 & 95\% Qu. & 10.99 & 9.98 & 0.92 & 1.00 & 0.68 & 1.21 & 1.02 \\
\end{tabular}
\vspace{0.4cm}
\end{center}

For a final comparison we consider a Support Vector Machine (SVM)
method \citep{drucker:1996} illustrated on this data and compared to
Bagging \citep{breiman:1996}.  We note that the SVM method required
cross-validation (CV) to set some of its parameters.  In the
comparison, 100 randomized training sets of size $n=200$ were used,
and RMSEs were collected for a (single) test set of size $n'=1000$.
An average MSE of 0.67 is reported, showing the SVM to be uniformly
better the Bagging method with an MSE of 2.26.  We repeated the
experiment for the GP LLM (which requires no CV!), and obtained an
average MSE of 0.293, which is 2.28 times better than the SVM, and
7.71 times better than Bagging.

\subsection{Boston housing data}
\label{sec:boston}

A commonly used dataset for validating multivariate models is the
Boston Housing Data \citep{harrison:78}, which contains 506 responses
over 13 covariates.  \citet{chip:geor:mccu:2002} showed
that their (Bayesian) treed LM gave lower RMSEs, on average, compared
to a number of popular techniques (the same ones listed in the
previous section).  Here we employed a treed GP LLM, which is a
generalization of their treed LM, retaining the original treed LM
as an accessible special case.  Though computationally more intensive
than the treed LM, the treed GP LLM gives impressive results.  To
mitigate some of the computational demands, the LLM can be used to
initialize the Markov Chain by breaking the larger dataset into
smaller partitions.  Before treed GP burn--in begins, the model is fit
using only the faster (limiting) treed LM model.  Once the treed
partitioning has stabilized, this fit is taken as the starting value
for a full MCMC exploration of the posterior for the treed GP LLM.
This initialization process allows us to fit GPs on smaller segments
of the data, reducing the size of matrices that need to be inverted
and greatly reducing computation time.  For the Boston Housing data we
use $(\gamma,\theta_1,\theta_2)=(10,0.2,0.95)$, which gives the LLM a
prior probability of $0.95^{13}\approx 0.51$, when the $d_i$'s are
large.

Experiments in the Bayesian treed LM paper \citep{chip:geor:mccu:2002}
consist of calculating RMSEs via 10-fold CV.  The data are randomly
partitioned into 10 groups, iteratively trained on 9/10 of the data,
and tested on the remaining 1/10.  This is repeated for 20 random
partitions, and boxplots are shown.  Note that the logarithm of the
response is used and that CV is only used to assess predictive error,
not to tune parameters.  Samples are gathered from the posterior
predictive distribution of the treed LM for six parameterizations
using 20 restarts of 4000 iterations.  This seems excessive, but we
followed suit for the treed GP LLM in order to obtain a fair
comparison.  Our ``boxplot'' for training and testing RMSEs are
summarized in the table below.  As before, linear regression (on the
log responses) is used for calibration.
\begin{center}
\singlespacing
\begin{tabular}{ll|rrrrrr}
        & & Min & 1st Qu. & Median & Mean & 3rd Qu. & Max \\
\hline
train & GP LLM  &0.0701& 0.0716& 0.0724& 0.0728& 0.0730& 0.0818 \\
& Linear        &0.1868 &0.1869 & 0.1869& 0.1869&0.1869 & 0.1870\\
\hline
test & GP LLM   &0.1321& 0.1327& 0.1346& 0.1346& 0.1356& 0.1389 \\
& Linear        &0.1926 &0.1945 & 0.1950& 0.1950&0.1953 & 0.1982
\end{tabular}
\vspace{0.4cm}
\end{center}
Notice that the RMSEs for the linear model have extremely low
variability.  This is similar to the results provided by Chipman et
al.~and was a key factor in determining that our experiment was
well--calibrated.  Upon comparison of the above numbers with the
boxplots in Chipman et al., it can readily be seen that the treed GP
LLM is leaps and bounds better than the treed LM, and {\em all} of the
other methods in the study.  Our worst training RMSE is almost two
times lower than the best ones from the boxplot.  All of our testing
RMSEs are lower than the lowest ones from the boxplot, and our median
RMSE (0.1346) is 1.26 times lower than the lowest median RMSE
($\approx 0.17$) from the boxplot.

More recently, \citet{chu:2004} performed a similar experiment (see
Table V), but instead of 10-fold CV, they randomly partitioned the
data 100 times into training/test sets of size 481/25 and reported
average MSEs on the un-transformed responses.  They compare their
Bayesian SVM regression algorithm (BSVR) to other high-powered
techniques like Ridge Regression, Relevance Vector Machine, GPs, etc.,
with and without ARD (automatic relevance determination---essentially,
a separable covariance function).  Repeating their experiment for the
treed GP LLM gave an average MSE of 6.96 compared to that of 6.99 for
the BSVR with ARD, making the two algorithms by far the best in the
comparison.  However, without ARD the MSE of BSVR was 12.34, 1.77
times higher than the treed GP LLM, and the worst in the comparison.
The reported results for a GP with (8.32) and without (9.13) ARD
showed the same effect, but to a lesser degree.  Perhaps not
surprisingly, the average MSEs do not tell the whole story.  The 1st,
median, and 3rd quartile MSEs we obtained for the treed GP LLM were
3.72, 5.32 and 8.48 respectively, showing that its distribution had a
heavy right--hand tail.  We take this as an indication that several
responses in the data are either misleading, noisy, or otherwise very
hard to predict.

%%%%%%%%%%% can we include the Neal results now too?
% The SVM method \citep{drucker:1996} was also used on this data
% obtaining an MSE of 7.2, and again compared this to Bagging which
% gave an MSE of 12.4.  Thus the treed GP LLM is $yy$ and $yyy$ times
% better than these two methods, respectively.

\section{Conclusions}
\label{sec:conclude}

Gaussian processes are a flexible modeling tool which can be an
overkill for many applications.  We have shown how its limiting linear
model can be both useful and accessible in terms of Bayesian posterior
estimation, and prediction.  The benefits include speed, parsimony,
and a relatively straightforward implementation of a semiparametric
model.  When combined with treed partitioning the GP LLM extends the
treed LM, resulting in a uniquely nonstationary, tractable, and highly
accurate regression tool.

We have focused on the separable power family of correlation
functions, but the methodology is by no means restricted to this
family.  All that is required is that the relevant family have a range
parameter (like $\mb{d}$) that yields the limiting (scaled) identity
covariance matrix characterizing the LLM (like $\mb{d} \rightarrow
0$).  For example, allowing a power $0 < p_i \ne 2$ in the power
family of Eq.~(\ref{e:cor_d}) is straightforward.  The separable
Mat\'ern family also has the desired property:
\[
K(\mb{x}_j, \mb{x}_k|\rho,\phi,\mb{d}) =
\prod_{i=1}^{m_X} 
\frac{\pi^{1/2}\phi d_i^{2\rho}}{2^{\rho-1}\Gamma(\rho+1/2)} 
(||x_{ij} - x_{ik}||/d_i)^\rho \mathcal{K}_\rho 
(||x_{ij} - x_{ik}||/d_i ),
\] 
where $\mathcal{K}_\rho$ is a modified Bessel function of the second
kind.  Unlike the power family, the isotropic Mat\'ern family does not
arise as a special case where $d_i = d$, for $i=1,\dots, m_X$, but, as
mentioned in Section~\ref{sec:model}, the LLM methodology remains
similarly applicable when there is only one range parameter.

One obvious extension of this work is to allow a larger class of
``simple'' models for the mean function of the process, such as
higher-order polynomials or interactions between linear terms.
Choosing among simpler models can be done straightforwardly within the
Bayesian framework
\citep{geor:mccu:1993,geweke:1996,jose:hung:sudj:2008}.  Such an
extension would likely allow more frequent selection of the simpler
model, reducing the need for the GP.  However, the clear understanding
of what limiting cases lead to jumping from a GP to a linear model, as
well as how to construct a set of booleans and their priors, would be
lost in moving beyond linear models.  Over a local region, a quadratic
function would be well--approximated by a GP with a range parameter
that could be neither very small nor large.  Thus the approach
contained herein would need further thought to efficiently extend it.
There is also the tradeoff of the extra computing resources needed to
test a larger set of models, whereas linear models can be worked
directly into the current model fitting (via the auxiliary booleans),
and do not require separate model selection steps.

We believe that a large contribution of the GP LLM will be in the
domain of sequential design of computer experiments \citep{gra:lee:2008}
which was the inspiration for much of the work presented here.
Empirical evidence suggests that many computer experiments are nearly
linear.  That is, either the response is linear in most of its input
dimensions, or the process is entirely linear in a subset of the input
domain.  Supremely relevant, but largely ignored in this paper, is
that the Bayesian treed GP LLM provides a {\em full} posterior
predictive distribution (particularly a nonstationary and thus
region--specific estimate of predictive variance) which can be used
towards active learning in the input domain.  Exploitation of these
characteristics should lead to an efficient framework for the adaptive
exploration of computer experiment parameter spaces.

%\small{
\bibliography{../btgpm/tgp}
\bibliographystyle{jasa}
%}

\end{document}